\documentclass[12pt]{spieman}  
\usepackage{amsmath, amsfonts, amssymb}
\usepackage{graphicx}
\usepackage{setspace}
\usepackage{tocloft}
\usepackage{lineno}

\newcommand{\apj}{ApJ}
\newcommand{\apjs}{ApJS}
\newcommand{\apjl}{ApJL}
\newcommand{\pasp}{PASP}
\newcommand{\mnras}{MNRAS}

\newcommand{\nat}{Nature}

\newcommand{\aap}{A\&A}

\newcommand{\araa}{ARA\&A}

\newcommand{\ssr}{Space Science Reviews}
\newcommand{\apss}{Ap\&SS}

\title{High-cadence observations of
galactic nuclei by the future
two-band UV-photometry mission
QUVIK}

\author[a*]{Michal Zaja\v{c}ek}
\author[a]{Norbert Werner}
\author[a]{Henry Best}
\author[a]{Jolie Esme L'Heureux}
\author[a]{Jakub \v{R}ípa}
\author[a]{Monika Pikhartová}
\author[a]{Martin Mondek}
\author[a]{Filip M\"{u}nz}
\author[a]{Lýdia Štofanová}
\author[a]{Petr Kurf\"{u}rst}
\author[a]{Matúš Labaj}
\author[a]{Izzy L. Garland}
\author[b]{Aaron Tohuvavohu}
\author[c]{Vladimír Karas}
\author[c]{Petra Suková}
\affil[a]{Department of Theoretical Physics and Astrophysics, Faculty of Science, Masaryk University,
Kotl\'{a}\v{r}sk\'{a} 2, 611 37 Brno, Czech Republic}
\affil[b]{Cahill Center for Astronomy and Astrophysics, California Institute of Technology, Pasadena, CA 91125, USA}
\affil[c]{Astronomical Institute of the Czech Academy of Sciences, Boční II 1401, 141 00 Prague, Czech Republic}

\cftpagenumbersoff{figure}
\cftpagenumbersoff{table} 
\begin{document} 
\maketitle

\begin{abstract}
The Quick Ultra-VIolet Kilonova surveyor (QUVIK), a two-band UV space telescope approved for funding as a Czech national science and technology mission, will focus on detecting early UV light of kilonovae (Werner et al., 2024). In addition, it will study the UV emission of stars and stellar systems (Krtička et al., 2024) as well as the intense and variable emission of active galactic nuclei (AGN) or galactic nuclei activated by tidal disruption events (Zajaček et al., 2024). In this contribution, we describe the role of this small ($\sim 30$-cm diameter) UV telescope for studying bright, nearby AGN. With its NUV and FUV bands, the telescope will perform high-cadence ($\sim 0.1$-1 day) two-band photometric monitoring of nearby AGN ($z<1$), which will allow us to probe accretion disk sizes/temperature profiles via photometric reverberation mapping. Thanks to its versatility, \emph{QUVIK} will be able to perform a moderately fast repointing ($<20$ min) to target candidates for tidal disruption events (TDEs). Early detection of the UV emission following a TDE optical flare, in combination with the subsequent two-band UV monitoring performed simultaneously with other observatories, will enable us to infer the time delay (or its lack of) between the optical, UV, and X-ray emission. In combination with theoretical models, it will be possible to shed more light on the origin of the UV/optical emission of TDEs. Furthermore, the two-band monitoring of nuclear transients will be beneficial in distinguishing between TDEs (nearly constant blue colour) and supernovae (progressive reddening) in the era of intensive wide-field surveys.    

\end{abstract}


{\noindent \footnotesize\textbf{*}\linkable{zajacek@physics.muni.cz} }


\section{Introduction}
\label{sect:intro}  
Detailed spectral and temporal studies of the ultraviolet (UV) domain of the electromagnetic spectrum ($\sim 100-400$ nm) are crucial for understanding the dynamics of stars, compact objects, accretion flows in both stellar systems and galactic nuclei, as well as for capturing the evolution of the most energetic phenomena in the Universe, such as $\gamma$-ray bursts due to collapsing massive stars or neutron-star binary mergers. Since the harder UV light is (fortunately) blocked by the Earth's atmosphere at an altitude of 30-40\,{\rm km} within the ozone layer, ground-based observations do not allow detection of a sufficient signal at shorter UV wavelengths, especially in the far-UV domain (122-200 nm). The development of UV astronomy is, therefore, ultimately linked to an ability to carry UV-sensitive detectors above the Earth's atmosphere (see Linsky 2018\cite{2018Ap&SS.363..101L} for a review).

The UV light consists of both continuum and line emission contributions. The continuum can be of thermal or non-thermal origin, which eventually determines the slope of the continuum spectral energy distribution (SED) in the UV domain or, in other words, the UV colour between far-UV (FUV) and near-UV (NUV) bands. Assuming optically thick, blackbody emission, in the FUV domain ($\sim 150$ nm) objects with the effective temperature of $T_{\rm NUV}\simeq b/\lambda=19320\,(b/0.29\,{\rm cm\,K}) (\lambda/150\,{\rm nm})^{-1}\,{\rm K}$ emit the most (Wien displacement's law), while in the NUV domain ($\sim 300$ nm) the effective temperature drops by a factor of two to $T_{\rm FUV}\sim 9660\,{\rm K}$. In terms of stellar spectral types, these temperatures correspond to the spectral types B ($\sim 10\,000-30\,000$ K) and A ($\sim 7\,300-10\,000$ K). In general, different types of celestial objects and associated processes can be studied in both NUV and FUV bands:
\begin{itemize}
    \item rapid stellar transients, such as the UV emission of kilonovae (depending on the explosion scenario), $\gamma$-ray burst afterglows, and supernovae of type I and II\cite{2020A&A...642A.214K,werner2024},
    \item hot and cool stars, their activity (stellar flares), stellar clusters, and the interstellar medium extinction curve \cite{krticka2024},
    \item active galactic nuclei (AGN) and nuclear transients, in particular tidal disruption events (TDEs--both full and partial, which can also be repeating), changing-look AGN, and high-cadence monitoring in different bands to perform reverberation mapping of the circumnuclear medium\cite{zajacek2024}. 
\end{itemize}

During the previous decades, there have been several space-borne UV telescopes that have brought ground-breaking results. The first UV space missions were carried out as a part of the \textit{Orbiting Astronomical Observatories} (OAOs) program between 1966 and 1981. Out of four OAOs, OAO-2 \textit{Stargazer} and OAO-3 \textit{Copernicus} conducted successful UV measurements. \textit{Stargazer} carried 11 UV telescopes with mirrors of 20 cm, which discovered large cometary haloes consisting mainly of hydrogen. It also led to the UV detection of novae whose UV emission brightened following the peak of the optical emission. \textit{Copernicus} (named to mark the 500th anniversary of Copernicus's birth) conducted joint UV and X-ray observations of selected objects. It collected UV spectra of hundreds of stars and discovered long-period X-ray pulsars with periods of several minutes. Concerning galactic nuclei, both \textit{Stargazer} and \textit{Copernicus} detected UV emission of 35 galaxies, among which were Seyfert galaxies NGC 4051 and NGC 1068 \cite{1972ARA&A..10..197B}. In Europe, pioneering UV and X-ray observations were performed by the \textit{TD-1} astrophysical research satellite launched in 1972 and by the \textit{Astronomical Netherlands Satellite} (\textit{ANS}) launched in 1974. 

Many significant measurements in the UV domain, including those related to the variability and spectral properties of AGN, were performed by the \textit{International Ultraviolet Explorer} (IUE), which was launched in 1978. The mission equipped with a mirror of 45 cm lasted for 18 years until 1996, and it was capable of short-wavelength (115-200 nm) and long-wavelength (185-330 nm) UV spectroscopy \cite{1978Natur.275..372B}. It provided several key measurements, including the studies of stellar winds and interstellar medium extinction, as well as detailed observations of the supernova SN 1987A, which deviated from standard stellar evolution models at the time. Concerning AGN, IUE performed detailed monitoring of the nearby bright Seyfert galaxy NGC 4151 ($z=0.0033)$, which exhibited variability on timescales of only several days \cite{1989ARA&A..27..397K}. More importantly, with respect to the continuum variability of NGC 4151, IUE detected time delays of several days for the UV broad lines (CIV, HeII, CIII], and MgII)\cite{2006ApJ...647..901M}, thus confirming the reprocessing scenario of the broad-line region (BLR) clouds that surround the supermassive black hole (SMBH) at a specific distance range that is traced by the broad lines of a certain ionization potential\cite{1988MNRAS.232..539C}. In combination with the detected line dispersion, it was possible to infer the virial SMBH mass of $M_{\bullet}=(4.14 \pm 0.74) \times 10^7\,M_{\odot}$ using $M_{\bullet}=f_{\sigma}c \tau_{\rm BLR} \sigma^2/G$ where $f_{\sigma}$ is the virial factor, $\tau_{\rm BLR}$ is the time delay of the line variability, and $\sigma$ is the line dispersion \cite{2006ApJ...647..901M}, with $c$ and $G$ as the speed of light and the gravitational constant. The monitoring of NGC 5548 ($z=	0.01651$) by IUE led to the determination of the spatial offset of the UV FeII/Balmer continuum complex (``small blue bump'') between 220 and 420 nm from the SMBH ($\sim 10$ days, which corresponds to $\sim 1\,700$ AU)\cite{1993ApJ...404..576M}. The UV FeII pseudocontinuum/Balmer continuum was found to be variable with an amplitude of $\sim 20\%$. The UV FeII emission, consisting of many FeII line transitions is an important contributor to the line cooling of the circumnuclear gas.  

More recently, a near- (175-275 nm) and far-UV (135-175 nm) survey was performed by the \textit{Galaxy Evolution Explorer} (\textit{GALEX})\cite{2005ApJ...619L...1M} satellite launched in 2003 and decommissioned in 2013. With its 50-cm mirror and a field of view of $1.2^{\circ}$, the all-sky survey reached the photometric sensitivity of $m_{\rm AB}\sim 20.5$ mag, with a deeper sensitivity in smaller areas of the sky. Its main focus was to track the star-formation history up to $z\sim 2$, i.e., during the last 10 billion years. \textit{GALEX} was also capable of performing slitless spectroscopy with a low resolution ($R\sim 100-200$) in the wavelength range of 135-275 nm. It played a crucial role in constraining the mean spectrum of type I quasars, alongside Spitzer NIR and MIR measurements, SDSS optical data, ROSAT X-ray measurements, and VLA radio data. In particular, wavelength-dependent bolometric corrections and their uncertainties were derived based on the quasar mean spectrum\cite{2006ApJS..166..470R}. \textit{GALEX} UV measurements have also been utilized to constrain SMBH masses, viewing angles, and even SMBH spins based on fitting type I AGN UV-optical spectra with the relativistic thin-disk model (see e.g. Modzelewska et al. 2014\cite{2014A&A...570A..53M} and Zaja\v{c}ek et al. 2020\cite{2020ApJ...896..146Z}).

Furthermore, important UV observations are currently provided by the \textit{Hubble Space Telescope} (HST) launched in 1990 and equipped with a 2.4-meter mirror, which directs the incoming photons into several near-UV and far-UV imagers and spectrographs, in particular the Advanced Camera for Surveys (ACS; near-UV imaging), Cosmic Origins Spectrograph (COS; UV spectra in the range 115-320 nm), Space Telescope Imaging Spectrograph (spatially resolved spectroscopy from 115 nm to 1030 nm, with the high spatial resolution echelle spectroscopy in the UV domain and photon-tagging to achieve a high temporal resolution), and Wide Field Camera 3 equipped with a UVIS channel for imaging between 200 and 1000 nm.

Another relevant instrument is the \textit{Neil Gehrels Swift Observatory} launched in 2004, which is equipped with an Ultra-Violet/Optical Telescope (UVOT) instrument onboard, in addition to $\gamma$-ray and X-ray sensitive detectors. Hence, \textit{Swift} is capable of intense multiwavelength monitoring of all kinds of transients, from $\gamma$-ray bursts to galactic nuclei transients. The UVOT telescope is a modified Ritchey-Chr\'etien with a 30-cm mirror, $2.5''$ point spread function (PSF; at 350 nm), and a sensitivity of 22.3 AB magnitude (in 1000 seconds in the B band). UVOT can perform imaging and photometry in six UV/optical filters (UVW2, UVM2, UVW1, U, B, V) from $170$ to $650$ nm \cite{2005SSRv..120...95R} within a 17'$\times$17' field. Concerning AGN research, both the HST and Swift provided high-cadence ($\sim 1$ day) UV light curves for accretion-disk reverberation-mapping (RM) of nearby type 1 AGN (for instance, NGC 5548 at $z=0.01651$\cite{2016ApJ...821...56F}). The interband time delays as a function of the band central wavelength are generally consistent with the standard accretion disk predictions for the reprocessing of the continuum emission, i.e. the relation $\tau(\lambda)=\alpha [(\lambda/\lambda_0)^{\beta}-1]$ with $\beta\simeq 4/3$ within the uncertainties, where $\tau$ is the time lag of the continuum emission at the wavelength $\lambda$ expressed with respect to the wavelength $\lambda_0$ and $\beta$ is the slope related to the temperature profile of the accretion disk, $T_{\rm disk}\propto r^{-1/\beta}$, and $\alpha$ is the normalization coefficient. Two basic discrepancies were found with the predictions of the standard accretion disk theory:
\begin{itemize}
  \item[(i)] the time delays/accretion-disk sizes tend to be larger by a factor of $\sim 3$,
  \item[(ii)] the $U$-band time delays are significantly larger than the best-fit time-delay/wavelength relation due to the contamination by the broad-line emission contribution (contamination by the Balmer continuum in the $U$ band reaches $\sim 20\%$)\cite{2021iSci...24j2557C}. In fact, the $U$-band excess is commonly detected. The HST spectroscopic monitoring of NGC 4593 showed that the feature is rather broad, extending from 300 to 400 nm, with a discontinuity at the Balmer jump (364.6 nm), which is consistent with the BLR contribution scenario \cite{2018ApJ...857...53C}.
\end{itemize}

Among other galactic nucleus discoveries, the \textit{Swift} Observatory target-of-opportunity monitoring recently led to the discovery of repeating nuclear transients with periods of about one month, which is intermediate between the X-ray quasi-periodic erupters (QPEs; from a few hours to a day) and the repeating partial TDEs (hundreds of days), see in particular the study of Swift J0230-28 with the periodicity of $\sim 22$ days\cite{2023NatAs...7.1368E,2024NatAs...8..347G,2024arXiv241105948P} and the ``Ansky'' source with the QPE period of $\sim 4.5$ days (ZTF19acnskyy\cite{2025arXiv250407169H,2025arXiv250407167C}).    

\textit{AstroSat} was launched in 2015 with the aim to carry out multiwavelength observations and imaging of variable sources on different scales, from stars and stellar binaries, to galaxies, galactic nuclei, and galaxy groups and clusters. It performs observations in the optical, UV, soft, and hard X-ray domains. Its Ultra-Violet Imaging Telescope (\textit{UVIT}) with a mirror size of $\sim 40$ cm can perform imaging with a field of view of $\sim 28'$ in diameter in three spectral channels, 130-180 nm, 180-300 nm, and 320-530 nm, with the possibility of performing a slitless grating low-resolution spectroscopy ($R\sim 100$). A relatively high angular resolution in the UV domain $\sim 1.8''$ allows the decomposition of a galaxy surface brightness into the host and the point-source AGN. \textit{AstroSat} observations in the X-ray/UV domain led to tighter constraints on broadband AGN spectra. In a few cases, relatively large accretion-disk inner radii of $\sim 35$-150 gravitational radii ($r_{\rm{g}}$) were found\cite{2021MNRAS.504.4015D,2023ApJ...950...90K}, which may correspond to the transition between the outer standard thin disk and the inner hot advection-dominated accretion flow\cite{2023MNRAS.522.2869S} for lower-accreting AGN. 

In the near future, time-domain UV astronomy will benefit significantly from the \textit{Ultraviolet Transient Astronomy
Satellite (ULTRASAT)} \cite{Ben-Ami2022,shvartzvald2023}, an Israeli space telescope with a 33 cm effective aperture and an unprecedentedly large field of view of 204 deg$^2$. \textit{ULTRASAT} is expected to be launched to a geostationary orbit at the end of 2027. It will perform a wide-field survey of variable and transient sources in the 230-290 nm band with a cadence of $\sim 2$ days, including supernovae, kilonovae, TDEs, and other transients.  

This contribution is structured as follows. We begin with Section~\ref{sec_QUVIK} where we introduce the concept of the Czech ambitious space telescope QUVIK. In Section~\ref{sect:primary}, we introduce the primary scientific objective of \textit{QUVIK} -- the early detection of the UV light of kilonovae. In Section~\ref{sect: AGN}, we feature one of the secondary scientific objectives -- AGN and (repeating) nuclear transients. In Section~\ref{sect: Discussion}, we briefly outline the synergy of \textit{QUVIK} monitoring and observations with other missions available at the time of its launch and active phase (2029-2032). We summarize the main characteristics of the \textit{QUVIK} mission in Section~\ref{sect: Summary}. 

\section{\textit{QUVIK} Space Telescope -- Czech Ambitious Space Mission}
\label{sec_QUVIK}

\begin{figure}
    \centering
    \includegraphics[width=\textwidth]{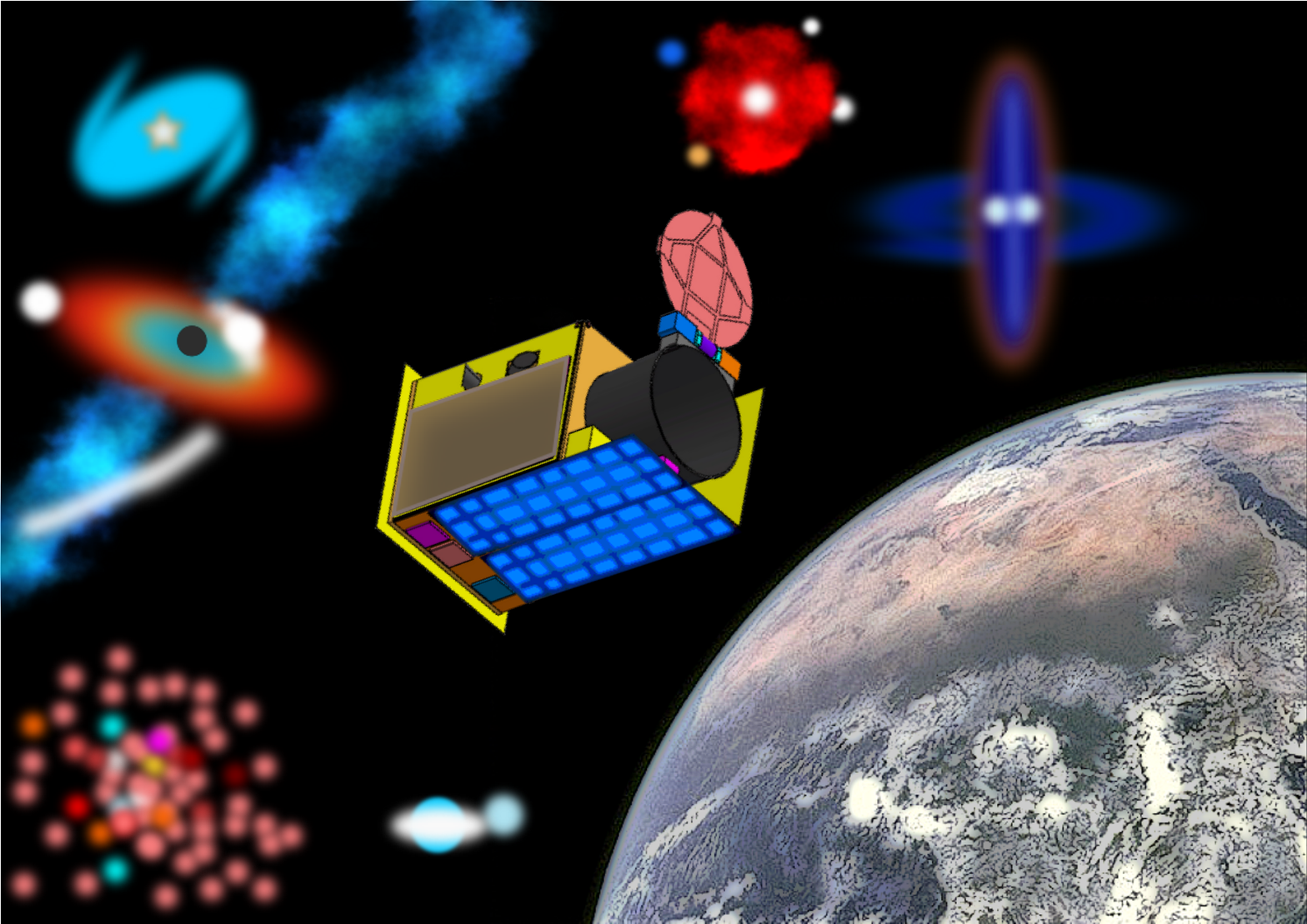}
    \caption{An artist's view of a two-band UV space telescope on a low-Earth orbit, which does not reflect the final detailed technical setup of \textit{QUVIK}. The accepted UV mission \textit{QUVIK} will carry two UV detectors, near-UV ($\sim 260-360$ nm) and far-UV ($\sim 140-190$ nm). The sketch also shows three main areas of the \textit{QUVIK} research: (i) stellar transients, such as kilonovae (primary research objective) and supernovae\cite{werner2024}, (ii) stars and stellar systems\cite{krticka2024}, (iii) active galactic nuclei and nuclear transients, such as tidal disruption events (TDEs)\cite{zajacek2024}.}
    \label{fig_artist_sketch}
\end{figure}

The \textit{Quick Ultra-VIolet Kilonova surveyor, QUVIK} mission~\cite{2022SPIE12181E..0BW} is designed to provide dedicated rapid follow-up observations of transients discovered by \textit{ULTRASAT} and other wide-field sky surveys \cite{werner2024,krticka2024,zajacek2024}. \textit{QUVIK} has been selected as a Czech national science and technology mission to be launched towards the end of this decade (2029-2030). The satellite will operate on a Sun-synchronous low-Earth orbit with an orbital period of $\sim90$ minutes. The primary mission shall last for three years. The mission will provide sensitive photometry in NUV and FUV bands complementary to the bandpass of \emph{ULTRASAT}. The mission baseline includes a NUV band of 260--360 nm and a FUV band of 140--190 nm \cite{werner2024}. Importantly, in the two UV bands, the mission will provide simultaneous photometry, which will also be important for the study of AGN \cite{zajacek2024}, in particular for the reverberation mapping of accretion flows. In addition, the simultaneous monitoring in the two UV bands will constrain the FUV-NUV colour evolution of transients, which will be crucial for their classification. In particular, it will help distinguish TDEs (constant blue colour) from supernovae (progressive reddening). In the NUV band, the field of view is aimed at $\gtrsim 1^{\circ}\times 1^{\circ}$ with a point spread function FWHM of $\lesssim 2.5$ arcseconds, which will allow isolating point sources (AGN) from the extended background emission (host galaxy).
The primary objective of \textit{QUVIK} is to perform a rapid follow-up of transient sources with the observation start latency of $\lesssim 20$ minutes after receiving a trigger from wide field-of-view surveys (e.g., \emph{ULTRASAT}, Legacy Survey of Space and Time by the Vera C. Rubin observatory) and gravitational-wave observatories detecting potential neutron star -- neutron star mergers (see Section~\ref{sect:primary} for more details). Furthermore, \textit{QUVIK} will be versatile enough to increase the monitoring cadence of transients according to the need, going potentially down to $\sim 0.1$ days for bright enough sources ($m_{\rm AB}\lesssim 20$ mag), for which the total integration time will be $\lesssim 10$ minutes. This will enable to capture short-term UV variability on the timescale of a few hours. In addition, complementary to the \textit{ULTRASAT} UV survey mission, \textit{QUVIK} will conduct high-cadence monitoring of selected targets in two UV bands, which will, in synergy with \textit{ULTRASAT}, result in constraints on the temporal evolution of the UV spectral slopes determined by the two-band flux density measurement by \textit{QUVIK} and the one-band measurement by \textit{ULTRASAT}.  In Fig.~\ref{fig_artist_sketch}, we show an artistic image of a two-band UV mission analogous to the \textit{QUVIK} featuring the main scientific aims of such a mission: stellar transients (kilonovae, supernovae, novae), stars and stellar populations, and AGN, including nuclear transients. \textit{QUVIK} will overall be constructed in a small-satellite design with the dimensions $70\times 70 \times 110$ cm and the projected weight of $\sim 200$ kg. It will be equipped with the near-real-time communication for the transient alert reception and the moderately fast repointing capability with the observation start latency of $\lesssim 20$ minutes\cite{werner2024}. Its primary payload will be the Cassegrain telescope with a mirror diameter of $\sim30$ cm. The NUV detector aims at a photometric sensitivity of 21.5 AB magnitude with a signal-to-noise ratio of 5 in one orbit. 
The overall simple and yet versatile concept of the \textit{QUVIK} is in line with the 2020 US Decadal survey, which highlighted the role of small- and medium-sized missions in advancing the time-domain and multi-messenger astrophysics\cite{2021pdaa.book.....N}.

\section{Primary scientific objective: Early detection of UV light of kilonovae}
\label{sect:primary}

The primary scientific objective of the \emph{QUVIK} mission is the early UV photometry of kilonovae. Kilonovae (KNe) originate from the mergers of binary neutron stars (BNS) or a neutron star and a black hole. In this process, an extensive amount of neutron-rich material is ejected. Within this dense ejecta, elements heavier than iron are created through the \textit{r-process}, i.e. rapid neutron-capture nucleosynthesis \cite{lattimer1974,wanajo2014}. The unstable \textit{r-process} elements decay and power the resulting kilonova transient \cite{li1998,kulkarni2005,metzger2010,metzger2019}. The name kilonova relates to the fact that the peak luminosity of the BNS merger is about a thousand times larger than a nova\cite{metzger2010}. On the other hand, a typical KN reaches about 1/100 to 1/10 of a typical supernova luminosity. The best-studied kilonova so far was AT2017gfo. It was produced by a BNS merger that was also the source of the gravitational-wave event GW170817 \cite{abbott2020a}. The extensive follow-up observations revealed that at the beginning, the emission was mainly in the NUV waveband. Later, over $\sim$10 days, it progressed through the blue and red optical emission to the infrared domain. The kilonova AT2017gfo was associated with the lenticular galaxy NGC 4993 ($z=0.009727$), located 140 million light years away.

The accompanying optical radiation of GW170817 was revealed 11 hours after the BNS merger and the UV radiation 4 hours afterwards \cite{abbott2017a,evans2017}. However, earlier observations are needed for a better understanding of the physics of KNe. The \emph{QUVIK} mission with its fast repointing capability ($\lesssim 20$ minutes) will provide early observations that can bring breakthroughs in our understanding of the physics of KNe. So far, we have sampled well multi-wavelength observations only for three KNe: GRB 160821B \cite{Lamb2019,Troja2019}, GW170817 / AT2017gfo \cite{abbott2017b}, and GRB 211211A \cite{Rastinejad2022}. Furthermore, only a few KN candidates associated with gamma-ray bursts (GRBs) have been observed: GRB 050709 \cite{Fox2005,Hjorth2005,Jin2016}, GRB 060614 \cite{GalYam2006, Yang2015}, and GRB 130603B \cite{Tanvir2013}. Therefore, the luminosity distribution of KNe is not well understood and more observations are required \cite{werner2024}.

The red and infrared emissions of the kilonova AT2017gfo were probably powered by the decay of \textit{r-process} elements. However, the origin of the UV and blue emissions dominating the first $\sim$1.5\,days is under discussion. It was suggested that the BNS coalescence did not collapse into a black hole immediately, but a short-lived magnetar was born leading to UV and blue emission \cite{metzger2018,curtis2023,combi2023}.

In general, the very early UV emission of a KN could also arise from shock interaction, a so-called cocoon emission \cite{Nakar2017}, or decay of free neutrons \cite{kulkarni2005}. It was suggested that the early-time power-law evolution of the luminosity of AT2017gfo could be explained by the cooling of shock-heated material around the neutron star merger \cite{Piro2018}. This heating was predicted to be produced as the GRB jet interacts with the merger debris, which is referred to as the cocoon emission \cite{Nakar2017,Gottlieb2018}. It was also predicted that free neutrons in the ejected material would decay via the $\beta$ process with a half-life of $\sim$15\,min,  delivering additional heating and intensifying the early KN emission \cite{kulkarni2005,metzger2015}. This would increase the NUV and FUV luminosity during the initial few hours \cite{metzger2019,Kulkarni2021}. The free neutron decay in a KN has not yet been confirmed. This is where the early photometry by \emph{QUVIK}, carried out within less than $\sim$6\,hours after the coalescence, would provide crucial information to constrain the KN UV emission mechanism.

Obtaining the early observations in NUV and FUV by \emph{QUVIK} will be important for distinguishing between different KNe models. Also, having simultaneous observations in UV and optical wavelengths would significantly constrain the models \cite{dorsman2023}. Due to the planned extensive ground-based follow-up efforts, additional optical and NIR observations will be available. For example, the ratios between the KN fluxes in UV, optical, and NIR wavelengths will help to constrain the properties of different ejecta, such as the ejecta mass, composition, thermal content, and the geometry of the ejecta \cite{metzger2019}.

Early multi-wavelength observations will also allow us to pinpoint the product of the merger. The following scenarios are possible: it can be a hyper-massive neutron star that rapidly collapses into a black hole; a rapidly spinning highly magnetised stable neutron star; or a BNS merger can directly collapse into a black hole. For each of these scenarios, different ratios of measured UV, optical, and NIR fluxes are predicted \cite{kasen2015,fernandez2016,bucciantini2012,yu2013,metzger2014}.

Moreover, some long GRBs can also be beneficial targets for \emph{QUVIK}. Although in most cases, long GRBs are known to come from the collapse of massive stars, for several long GRBs, despite deep optical observations, an accompanying supernova was not discovered (e.g., GRB 060605 and GRB 060614) \cite{Fynbo2006}. Furthermore, a special class of short GRBs with soft extended emission (EE-SGRBs), which by its duration resemble long GRBs, were suggested to arise from the coalescence of compact objects \cite{Norris2002, Norris2006, Gehrels2006}. Two recent observations seem to confirm this conjecture. The observations of GRB 211211A with the prompt gamma-ray duration of more than 30\,s pinpointed a kilonova \cite{Rastinejad2022, Troja2022} and the observations of the exceptionally bright long-duration GRB 230307A also revealed a kilonova \cite{2024Natur.626..737L}. Therefore, targeting long-duration GRBs with \emph{QUVIK} may, in some cases, unveil new KNe as well.

\section{Exemplary secondary objective:  Active galactic nuclei and nuclear transients}
\label{sect: AGN}

The \emph{QUVIK} mission will have the ability to quickly observe targets of opportunity and provide rapid observations in NUV and FUV bands.
This is ideal to study many types of transient events beyond the primary scientific objective, such as TDEs, AGN flaring events, changing look AGN (CL-AGN), and microlensing high-magnification events (HMEs). 
Furthermore, it offers the possibility to perform reverberation mapping of the AGN's inner accretion disk due to the NUV and FUV bands.

\begin{figure}
    \centering
    \includegraphics[width=0.8\textwidth]{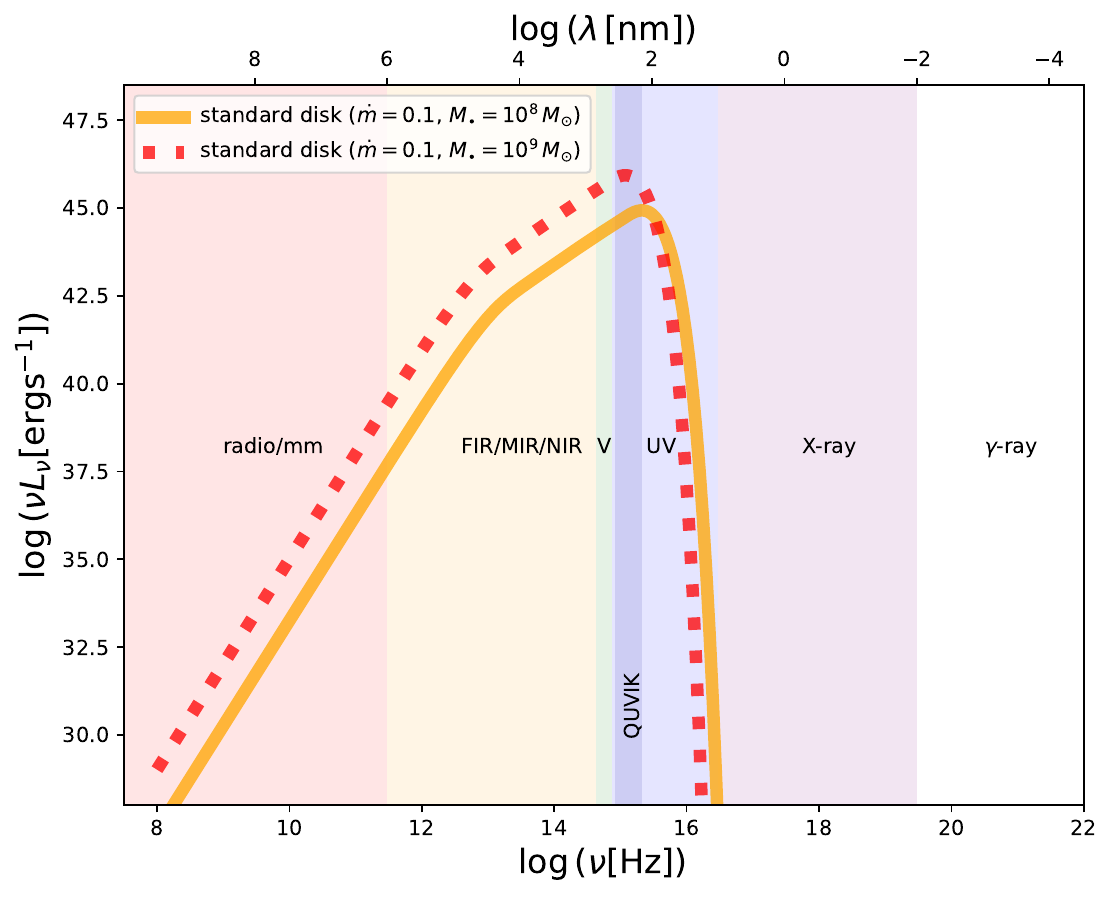}
    \caption{Spectral energy distributions (SEDs) of standard thin accretion disks (expressed in luminosities in ${\rm erg\,s^{-1}}$) around supermassive black holes with the masses of $M_{\bullet}=10^8\,M_{\odot}$ (orange solid line) and $M_{\bullet}=10^9\,M_{\odot}$ (red dotted line). In both cases the relative accretion rate (Eddington ratio) is set to $\dot{m}=0.1$ and the inclination is set to $\iota=20^{\circ}$. The vertical coloured bands correspond to different SED waveband ranges (from radio/mm to $\gamma$-ray). The dark-blue band denotes the \textit{QUVIK} FUV- and NUV-band range between 140 and 360 nm.}
    \label{fig_disk_SED}
\end{figure}

\begin{figure}
    \centering
    \includegraphics[width=0.8\textwidth]{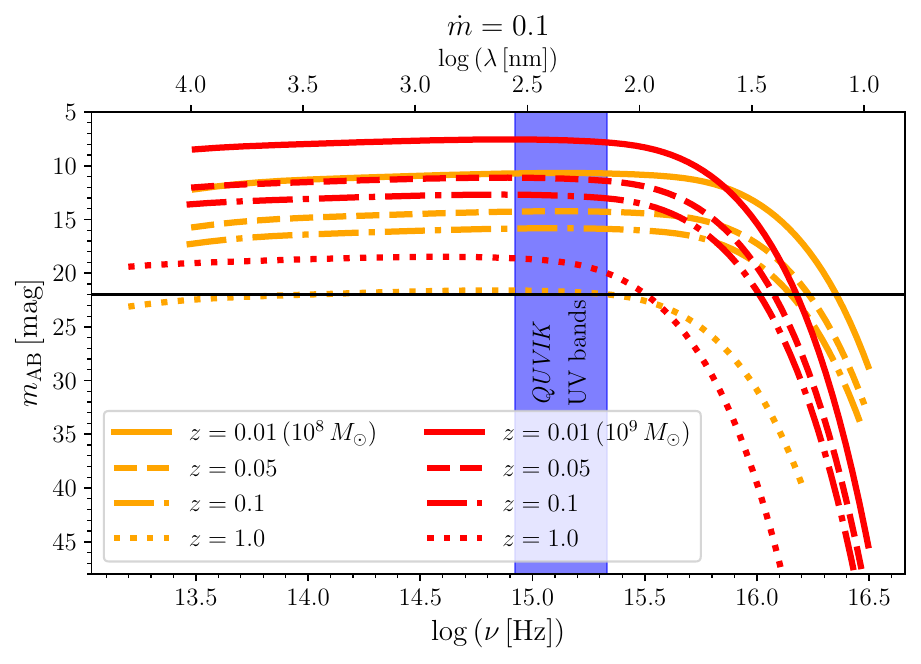}
    \caption{Apparent magnitudes of the emission of the standard thin disk (expressed in AB magnitudes) surrounding the SMBHs of $M_{\bullet}=10^8\,M_{\odot}$ (orange lines) and $M_{\bullet}=10^9\,M_{\odot}$ (red lines) as a function of the frequency (lower axis) and wavelength (upper axis). Different lines correspond to the sources at different redshifts -- $z=0.01$ (solid lines), $z=0.05$ (dashed lines), $z=0.1$ (dash-dotted lines), and $z=1.0$ (dotted lines) -- for both SMBH masses. The relative accretion rate is set to $\dot{m}=0.1$ and the inclination to $\iota=20^{\circ}$. The horizontal black line corresponds to the projected sensitivity limit of 22 AB magnitudes for the \textit{QUVIK} mission. The blue-shaded area corresponds to the anticipated far-UV and near-UV \textit{QUVIK} waveband range between $140$ and $360$ nm.}
    \label{fig_disk_mAB}
\end{figure}

For AGN, for which we can directly observe the emission of the accretion disk (type I when we look close to the symmetry axis of the system), the primary contributor to the UV emission in the AGN spectra is the thermal emission of the accretion disk itself. The excess emission in the UV domain is referred to as the ``big blue bump'' \cite{1987ApJ...321..305C}. For many AGN, it can be approximated by the thermal emission of a stationary standard thin accretion disk (classical Shakura-Sunyaev\cite{1973A&A....24..337S} or relativistic Novikov-Thorne\cite{1973blho.conf..343N} solutions), taking into account additional opacity and inclination effects \cite{1987ApJ...321..305C}. The most luminous AGN -- quasars -- typically accrete close to the upper limit when the radiative force acting on electrons via Thomson scattering nearly balances the gravitational force acting mostly on protons in bound Coulomb pairs of electrons and protons (Eddington limit). The corresponding Eddington accretion rate $\dot{M}_{\rm Edd}$ can be inferred from the maximum Eddington luminosity $L_{\rm Edd}$ taking into account the radiative efficiency $\eta_{\rm rad}$, which we set to $\sim 0.1$ (10\%),
\begin{align}
    \dot{M}_{\rm Edd}&=\frac{L_{\rm Edd}}{\eta_{\rm rad}c^2}=\frac{4\pi G M_{\bullet} m_{\rm p}}{\eta_{\rm rad} \sigma_{\rm T} c}\,\notag\\
    &\simeq 2.2 \left(\frac{M_{\bullet}}{10^8\,M_{\odot}} \right) \left(\frac{\eta_{\rm rad}}{0.1} \right)^{-1} {\rm M_{\odot}\,yr^{-1}}\,,
\end{align}
where $M_{\bullet}$ is the SMBH mass, $G$ is the gravitational constant, $m_{\rm p}$ is the proton mass, $\sigma_{\rm T}$ is the Thomson scattering cross-section, and $c$ is the speed of light. Thus quasars typically accrete of the order of one Solar mass per year. It is convenient to express the accretion rate of AGN relative to the Eddington rate. Assuming the same radiative efficiency, this is referred to as the relative accretion rate, $\dot{m}\equiv \dot{M}/\dot{M}_{\rm Edd}$, or the Eddington ratio (often denoted as $\lambda_{\rm Edd}=\dot{m}$). The standard thin-disk spectrum is given as the sum of black-body spectra of thin rings with temperatures following the power-law profile $T_{\rm disk}\propto r^{-3/4}$ as a function of radius. In Fig.~\ref{fig_disk_SED}, we show example spectral energy distributions (SEDs) of standard accretion disks surrounding the SMBHs of $M_{\bullet}=10^8\,M_{\odot}$ and $10^9\,M_{\odot}$, whose relative accretion rates are $\dot{m}=0.1$ and they are viewed at $\iota=20^{\circ}$. We see the basic trend that SMBHs with higher masses have larger UV luminosities and their SEDs are softer, i.e. the SED peak is shifted towards the NUV waveband. When unobscured by intervening circumnuclear gas and warm dust, AGN of type I can be detected by the \textit{QUVIK} UV mission across a wide redshift range from the local universe to the intermediate redshifts around $z\sim 1$, see Fig.~\ref{fig_disk_mAB}. This is especially the case of AGN with SMBHs of a higher mass of $M_{\bullet}=10^8-10^9\,M_{\odot}$. For a relative accretion rate of $\dot{m}=0.1$, the inclination of $\iota=20^{\circ}$, and the SMBH mass of $M_{\bullet}=10^8\,M_{\odot}$, the estimated UV AB magnitude is $m_{\rm AB}=10.7$ mag at $\lambda=310$ nm and $z=0.01$. At $z=1$, it is $m_{\rm AB}=21.6$ mag, which is at the sensitivity limit of the \textit{QUVIK} in the NUV band. However, for larger SMBH masses and higher relative accretion rates, the AGN brightness in the UV domain increases, hence the prospects for the detection of type I AGN and quasars at low- to intermediate-redshift are favourable (see Fig.~\ref{fig_disk_mAB} for the comparison between $M_{\bullet}=10^8\,M_{\odot}$ and $M_{\bullet}=10^9\,M_{\odot}$ at different redshifts). 

In the following, we focus on reverberation mapping of accretion disks in the UV domain (Subsection~\ref{subsec_reverberation}), stochastic nuclear transients (Subsection~\ref{subsec_stochastic}), and repeating nuclear transients (Subsection~\ref{subsec_repeating}) for simplicity, though some of the observed phenomena may combine these aspects and outlined observational methodologies.  

\begin{figure}
    \centering
    \includegraphics[width=\textwidth]{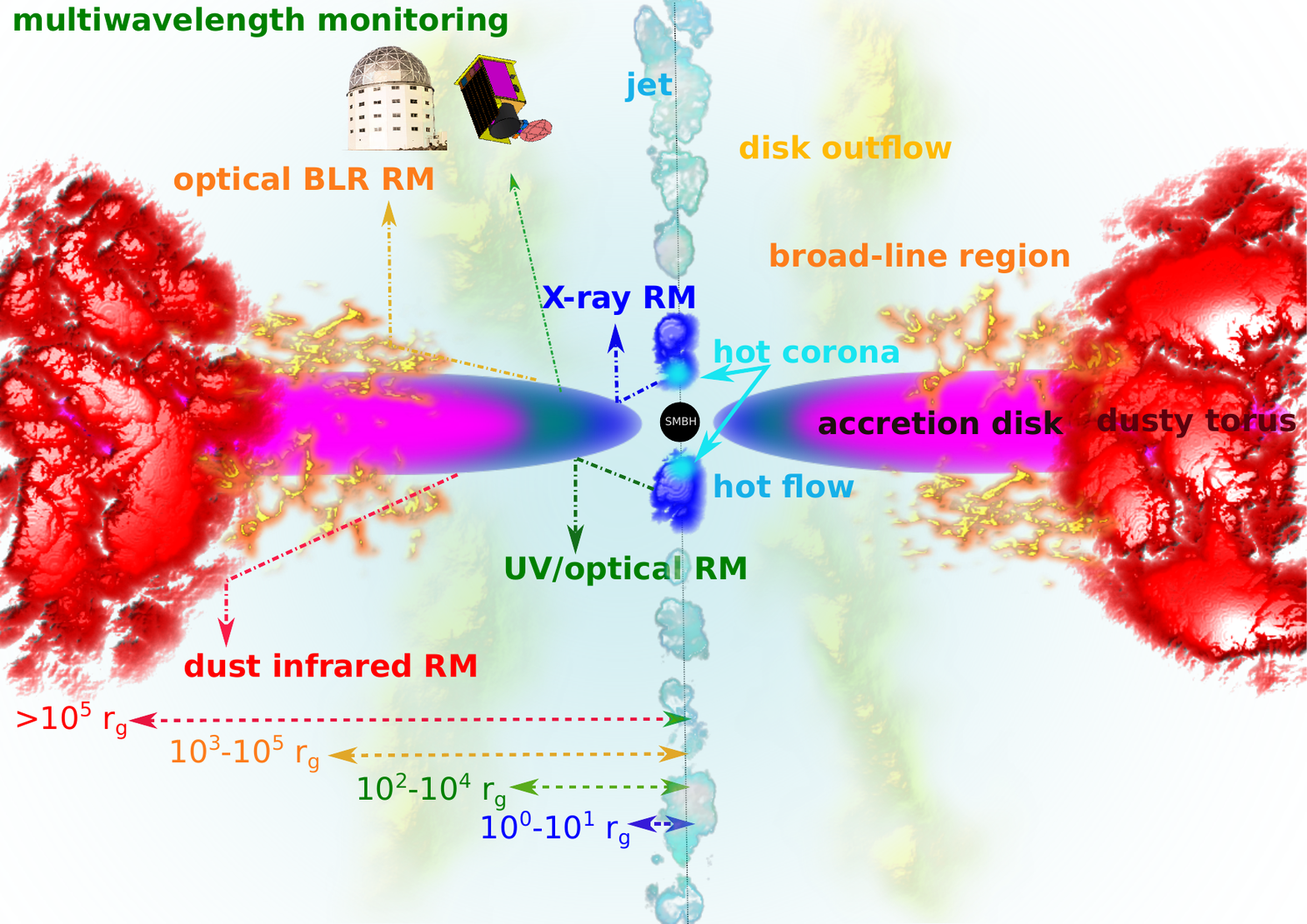}
    \caption{Illustration of different components of the unified AGN model. The figure depicts the hot corona, accretion disk, broad-line region (BLR), dusty torus, and how they are related to different spectral parts in the continuum and the line emission of AGN. The lower part of the image shows the typical length-scales associated with different wavebands ranging from the hard, soft X-ray, UV/optical, to the infrared domain.}
    \label{fig_AGN_unified}
\end{figure}

\subsection{Reverberation mapping of accretion disks}
\label{subsec_reverberation}

\begin{figure}
    \centering
    \includegraphics[width=0.9\linewidth]{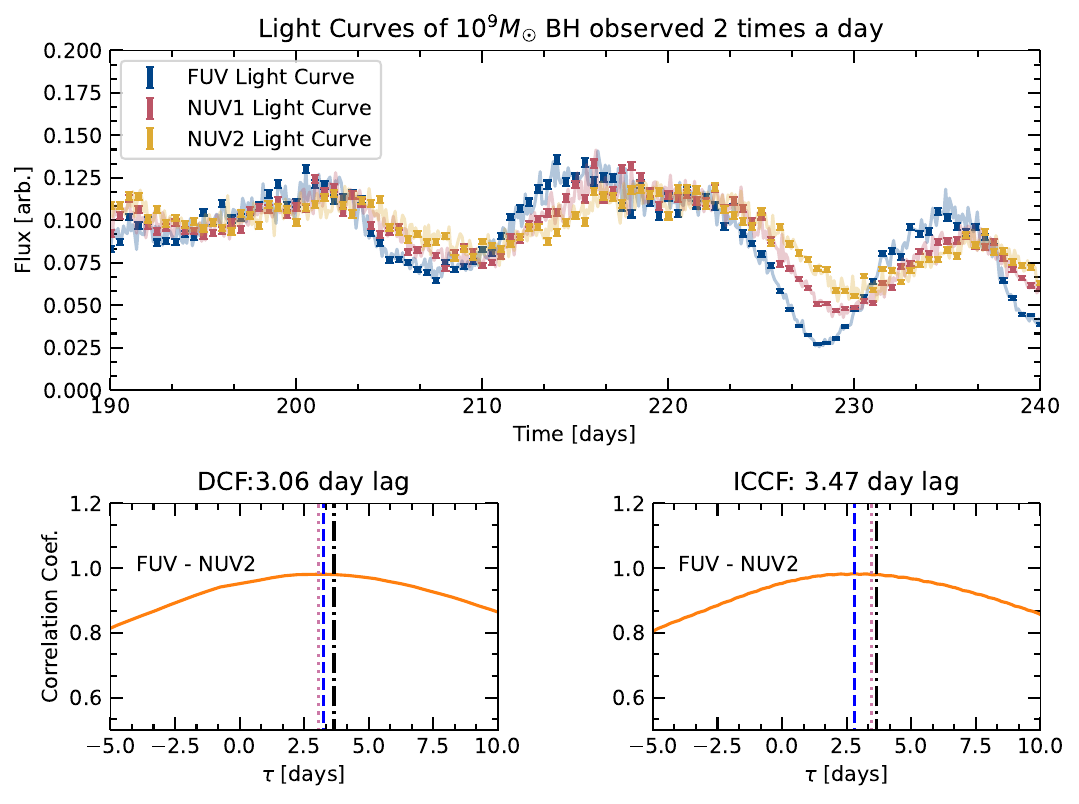}
    \caption{Top panel: Simulated light curves of a reverberating accretion disk for $M_{\bullet} = 10^{9} M_{\odot}$ as modeled in the FUV (175 nm), NUV1 (260 nm), and NUV2 (325 nm) bands. Dense light curves are plotted in faded colors, while dark points represent mock observations with a cadence of 0.5 days. Lower-left panel: Discrete correlation coefficient as a function of the time lag, $\tau$, prior to bootstrapping between the FUV and NUV2 bands. Lower-right panel: Interpolated cross-correlation coefficient as a function of $\tau$ between the FUV and NUV2 bands. Within the lower panels, the blue vertical dashed line represents the peak correlation coefficient, while the pink vertical dotted line represents the centroid measured at 80 \% of the peak. The black vertical dash-doted line represents the mean time delay as calculated directly from the means of the transfer functions.}
    \label{fig:light_curves}
\end{figure}

\begin{figure}
    \centering
    \includegraphics[width=0.9\linewidth]{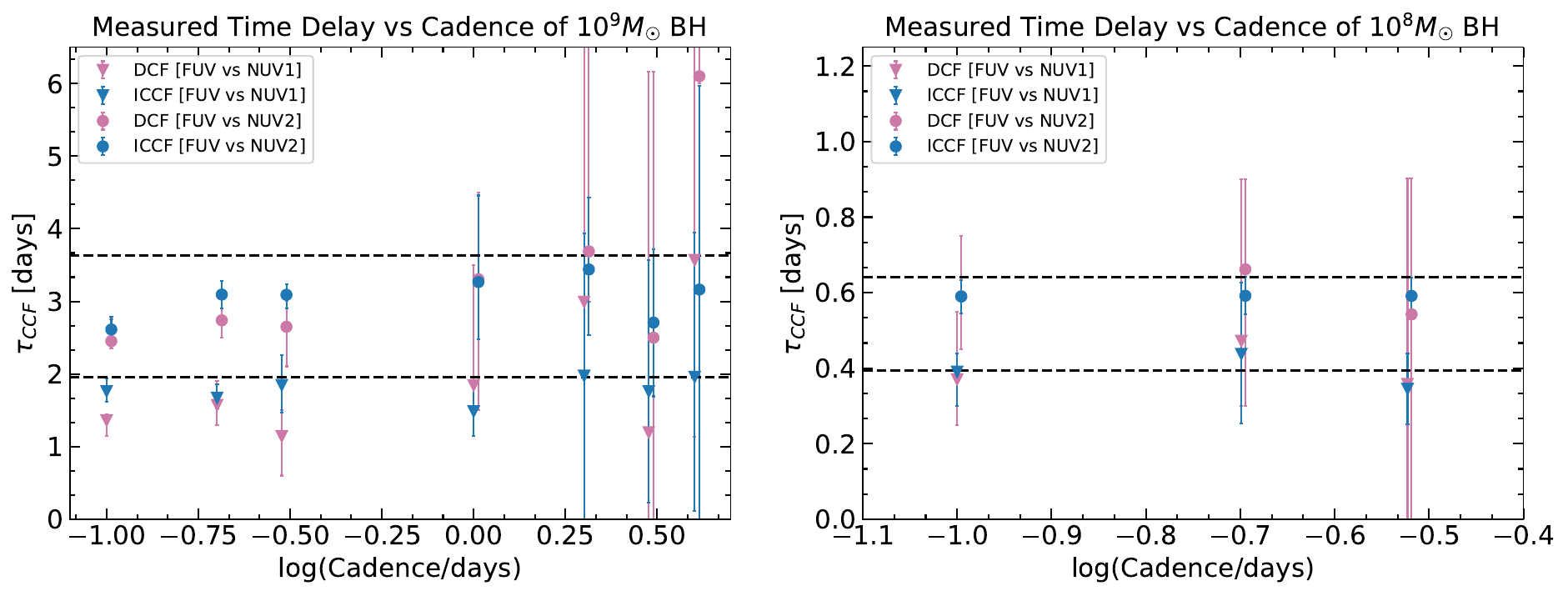}
    \caption{Measured time delays between FUV (175 nm), NUV1 (260 nm), and NUV2 (325 nm) as a function of observational cadence for both $M_{\bullet} = 10^{9}, 10^{8} M_{\odot}$ presented in the left and right panels, respectively. Error bars represent a 1 $\sigma$ uncertainty. Horizontal black dashed lines represent mean time delays for NUV1 and NUV2 with respect to the FUV band, as calculated from the means of the transfer functions. The cadence coordinates of the NUV2 bands are artificially shifted by  0.3 \% for $M_{\bullet} = 10^{9} M_{\odot}$ and 0.1 \%  for $M_{\bullet}  = 10^{8} M_{\odot}$ for clarity.}
    \label{fig:time_lags}
\end{figure}

With the two UV bands, \emph{QUVIK} has the ability to perform reverberation mapping in UV wavelengths.
The AGN's UV emission is likely due to a relatively compact region of the accretion disk near the supermassive black hole (SMBH)~\cite{Kaspi18, Cackett21, Kovacevic22}.
Spectra of accretion disks are more complex than a simple black body emission of typical accretion disk models (e.g. thin disk~\cite{ShakuraSunyaev, PageThorne74}, slim disk~\cite{Wang99}, and advection dominated accretion flows or ADAFs~\cite{Abramowicz13}).
Magneto-hydrodynamic (MHD) simulations tend to describe the accretion flow as composed of several gaseous components with different temperatures, densities, and geometric properties ~\cite{Evans88,2019ApJ...884L..37L,2022MNRAS.514..780W,Kaaz23, Secunda24}.
The real accretion disk is likely a combination of many of these accretion flows at different size scales.
While optical continuum reverberation mapping seems to support some aspects of these accretion flow models (albeit with $\sim$2 to 3 times larger accretion disk sizes~\cite{2016ApJ...821...56F,Jiang17}), there are fewer studies of UV reverberation~\cite{Kaspi18,2023MNRAS.521.4109K,2024MNRAS.527.5668K,2024A&A...691A..60P}. The contribution of UV photometric and spectroscopic observations to the multiwavelength monitoring of AGN has been essential in constraining accretion-disk temperature profiles, sizes, and the geometry and the kinematics of the circumnuclear medium; see e.g. the \textit{Swift} UVOT contribution discussed in Section~\ref{sect:intro}, the AGN STORM campaign on NGC 5548\cite{2015ApJ...806..128D}, and the AGN STORM 2 campaign on Mrk 817\cite{2021ApJ...922..151K}.   

\emph{QUVIK} will help to fill the gap by enabling the observations of a larger
sample of sources more easily, especially by employing the higher cadence required to probe the smaller size scales of the inner accretion disk.
In Section~\ref{sect: Discussion}, we describe how this can be joined with other surveys at different wavelengths (optical and X-ray) to provide reverberation mapping across a wider spectrum.

Here, we perform simulations of AGN light curves in order to test the efficacy of \emph{QUVIK} on reverberation mapping utilizing modular code \texttt{Amoeba}~\cite{Best24b}.
The UV emission of the AGN is simulated as the reprocessing of a stochastic driving signal within the lamppost model, where a point-like corona irradiates the accretion disk from a position directly above the black hole~\cite{Sergeev05, Cackett07}; see also Fig.~\ref{fig_AGN_unified} for the illustration of the AGN setup that is inspired by the unified AGN model\cite{1995PASP..107..803U}. 
We generate light curves assuming a Schwarzschild SMBH with masses of $M_{\bullet} = 10^{8}, 10^{9} M_{\odot}$, accreting with an Eddington ratio of 10 \% and placed at a redshift of 0.03.
The corona height is taken to be 6 $r_{\rm{g}}$, which is equal to the inner radius of the accretion disk, and it is assumed that it radiates with an average power of 10 \% of the bolometric luminosity. We then model the driving signal as a broken power-law (BPL) function, which has a characteristic time scale ($T_{B}$) based off of equation $\log T_B = 2.1 \log (M_{\bullet}/10^{6}M_{\odot}) - 0.98 \log( L_{\mathrm{Bol}}/10^{44}\mathrm{ergs~s}^{-1})  - 2.32$ \cite{mchardy2006}, emphasizing that this driving signal is not an observable quantity. The power spectral density (PSD) is taken to be $\propto f^{-1}$ for low frequencies, and $\propto f^{-2}$ for high frequencies. This corresponds to a break-point timescale of 86 and 7 days for $M_{\bullet} = 10^{9}$ and $10^{8} M_{\odot}$, respectively.

From this simulation, the UV signals are computed assuming that the accretion disk thermally responds when the driving signal is absorbed.
We computed this response by calculating the transfer functions of the accretion disk when viewed at 20$^{\circ}$\cite{Cackett07}.
These transfer functions encode the geometry and physics of the models used.
It is known that inclined accretion disks do not have a significant change in their mean response function, but can change the skew of the transfer function.
The UV signals are then the convolution of these transfer functions and the driving signal, to which we add 2 \% Poisson noise and sample the dense light curves based on various observation strategies.
We perform simulations for 3 UV bands: FUV, NUV1, and NUV2 using effective wavelengths of 175, 260, and 325 nm, respectively, where FUV and NUV2 correspond to the currently preferred \textit{QUVIK} bands and NUV1 corresponds to the \textit{ULTRASAT} band. 

To compute the time lag between the UV light curves, we then use two methods known as the Discrete Correlation Function (DCF)~\cite{edelson1988} and the Interpolated Cross-Correlation Function (ICCF)~\cite{peterson1998, sun2018}. These methods compute the correlation coefficient based on the time delay between two signals, where 1.0 represents perfect correlation.
The time lag between two signals may be taken to be the time lag associated with the peak correlation coefficient, or alternatively, the centroid value at a specified level~\cite{Sergeev05}.
Common thresholds for the centroid value are 50 or 80 \% of the peak correlation coefficient, which is found to give more reliable time lag measurements with respect to the peak~\cite{Derosa15, Korista19}.
In doing so, we are less biased towards how sharp or skewed the transfer functions are.
The uncertainties in the time lag are then computed using a bootstrapping method for the DCF and a Markov Chain Monte Carlo method for the ICCF, which improves our understanding of the time lag between the UV bands. For these mock observations, we set the UV light curve lengths to 180-day intervals.

While this model setup is very specific and focused on nearby, bright sources, its aim is to qualitatively explore the parameter space of possible cadences to assess the feasibility of continuum reverberation mapping in the UV domain, i.e. the measurement of the time lag between the FUV and NUV bands. In this regard, 2\% uncertainty is also fixed and assumes that the AGN sources are detected with \textit{QUVIK} at the signal-to-noise ratio of 100 and higher. At the moment, we do not consider a specific \textit{QUVIK} UV band throughput, which will be available in the future. In addition, we also do not consider intermediate-redshift sources, for which the intrinsic rest-frame time lag $\tau_0$ is noticeably prolonged by the factor $(1+z)$ in the observer's frame, which is negligible for the sources at $z<0.1$ considered here ($z=0.03$). The focus on nearby bright sources stems from the projected sensitivity of QUVIK FUV and NUV instruments, down to $m_{\rm AB}\sim 21-22$ mag, see also Fig.~\ref{fig_disk_mAB} for the analysis of the dependency of the apparent brightness on the Type I AGN redshift and the SMBH mass. 

The top panel of Fig.~\ref{fig:light_curves} illustrates the time lag between the FUV and NUV1 bands for an accretion disk associated with $M_{\bullet} = 10^{9} M_{\odot}$. 
The dense light curves are plotted for illustrative purposes, and the mock observations are presented as the dark points with error bars.
The time lag is clearly visible in this case, though there is also some smoothing of the signal which is representative of the differential accretion disk sizes.
The lower-left and lower-right panels represent the peak and centroid time delays computed using the DCF and ICCF methods, respectively.
These are plotted alongside the expected time delays as measured from the means of the transfer functions in vertical black dashed lines.
The skewed nature of the transfer functions can lead to an underestimation of the cross correlation measurements with respect to the expected mean value, so we expect an underestimation of approximately 20 \% using these methods~\cite{Chan20}.

The measured time lag uncertainties are then computed from the $\tau$ distributions, and are then presented in Fig.~\ref{fig:time_lags}.
For the DCF method, the time lag distribution is bootstrapped for 10,000 iterations to estimate the uncertainty of the time delay measurement, while the ICCF method is iterated with a Markov Chain Monte Carlo simulation for 10,000 iterations. 
We explore multiple cadences which could be realized with the \emph{QUVIK} telescope, which has a strong effect on the measured time lag and uncertainty.
Fig.~\ref{fig:time_lags} highlights that for $M_{\bullet} = 10^{9} M_{\odot}$, the time lags between the NUV, FUV1, and FUV2 bands may be realized with daily cadence. 
Tighter constraints are found at higher cadences, showcasing the need for multiple observations a day in order to study the inner accretion disk structure.
This can be complimented with optical and X-ray observations to extend the coverage, though other challenges arise in these wavelength ranges.
We note that multiple observations per day are required in order to get meaningful time delays between the FUV, NUV1, and NUV2 bands for more typical AGN with $M_{\bullet} = 10^{8} M_{\odot}$.\\
\textit{\underline{QUVIK observational strategy:}} For the reverberation mapping in the UV bands, \textit{QUVIK} can aim for AGN sources at the redshift of $z<1$, see also Fig.~\ref{fig_disk_mAB}. The higher the redshift of the source, the more blueward part of the spectrum in the rest frame \textit{QUVIK} can probe. In terms of a number of available sources,  within $z<0.5$, there are $\sim 151$ sources with a limiting magnitude of 17 in the SDSS $u$ band (Data Release 7)\cite{zajacek2024}. For these sources, the total integration time with \textit{QUVIK} to reach the signal-to-noise ratio of 100 is $\sim 100$ seconds\cite{zajacek2024}. Based on the analysis of mock observations with varied cadence and light-curve length, Zaja\v{c}ek et al.\cite{zajacek2024} concluded that the AGN with lower SMBH masses ($\sim 10^7\,M_{\odot}$) need to be monitored with a high cadence of $\sim 0.1$ days but a relatively short duration of the order of $\sim 10-100$ days is overall sufficient to capture the significant time delay between FUV and NUV bands, depending on the SMBH mass. From the past monitoring campaigns of nearby AGN, a monitoring lasting for about half a year is expected to be sufficient (see e.g. the monitoring of NGC5548 \cite{2016ApJ...821...56F} with $M_{\bullet}=5\times 10^7\,M_{\odot}$; Mrk 817\cite{2021ApJ...922..151K} with $M_{\bullet}=3\times 10^7\,M_{\odot}$), while the duration can decrease to a few 10 days for low SMBH-mass systems (see e.g. NGC4593\cite{2018ApJ...857...53C} with $M_{\bullet}=8\times 10^6\,M_{\odot}$, for which the campaign lasted 26 days). On the other hand, systems with larger SMBH masses ($\sim 10^8\,M_{\odot}$) can be monitored with a lower cadence of $\sim 1$ day, but a longer monitoring time of at least $\sim 180$ days is required. This is consistent with the analysis presented here where the light-curve length was fixed to 180 days. However, in case more UV bands are involved in the campaign (FUV, NUV1, and NUV2), a higher-cadence monitoring at $~\sim 0.1$ days is needed to detect significant time lags between the individual UV bands (FUV-NUV1 vs. FUV-NUV2).

\subsection{Stochastic nuclear transients}
\label{subsec_stochastic}

\begin{figure}[h!]
    \centering
    \includegraphics[width=\textwidth]{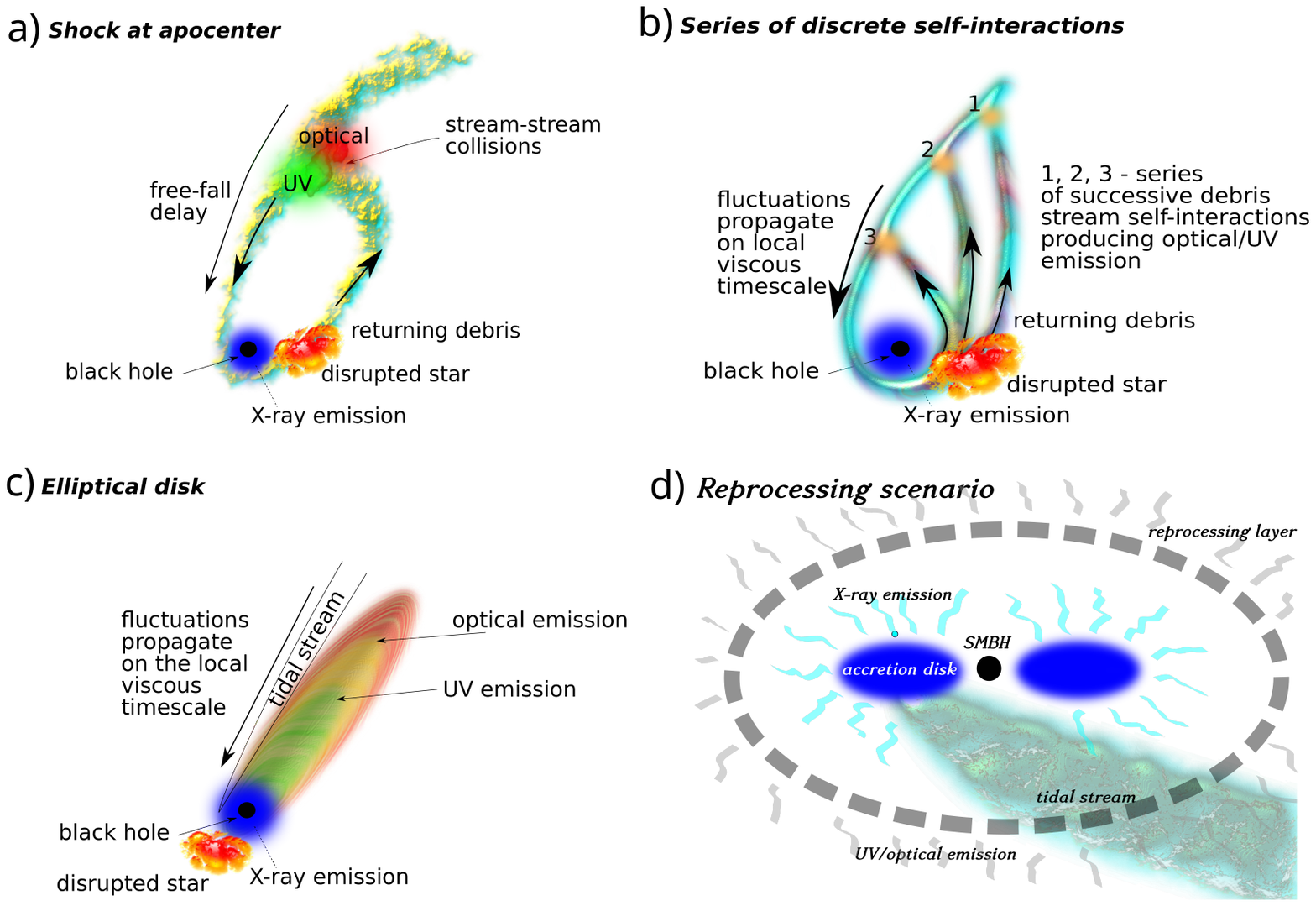}
    \caption{Different scenarios for the production of UV, optical, and X-ray emission in TDEs. a) Shock-at-apocenter scenario when the TDE stream collides with itself, gets shocked, and produces UV/optical emission. Fluctuations due to steam-stream collisions propagate inwards on the free-fall timescale. UV/optical light curve variability leads the X-ray emission. b) UV/optical emission is produced in the series of discrete stream-stream collisions that propagate towards the inner flow on the local viscous timescale. The X-ray emission follows the UV/optical emission. c) UV, optical, and the X-ray emission correspond to different radii of the elliptical accretion disk. Fluctuations propagate inwards on the local viscous timescale, which is significantly shorter than for the circular disk of a similar length-scale. In this scenario, the UV/optical variability leads the X-ray light curve. d) Reprocessing scenario, in which the variable X-ray emission generated within the inner accretion disk is reprocessed within the reprocessing layer at larger radii and re-radiated as a UV/optical emission. In this model, the X-ray emission leads the UV/optical emission. }
    \label{fig_TDE_models}
\end{figure}

\begin{figure}[h!]
    \centering
    \includegraphics[width=0.8\textwidth]{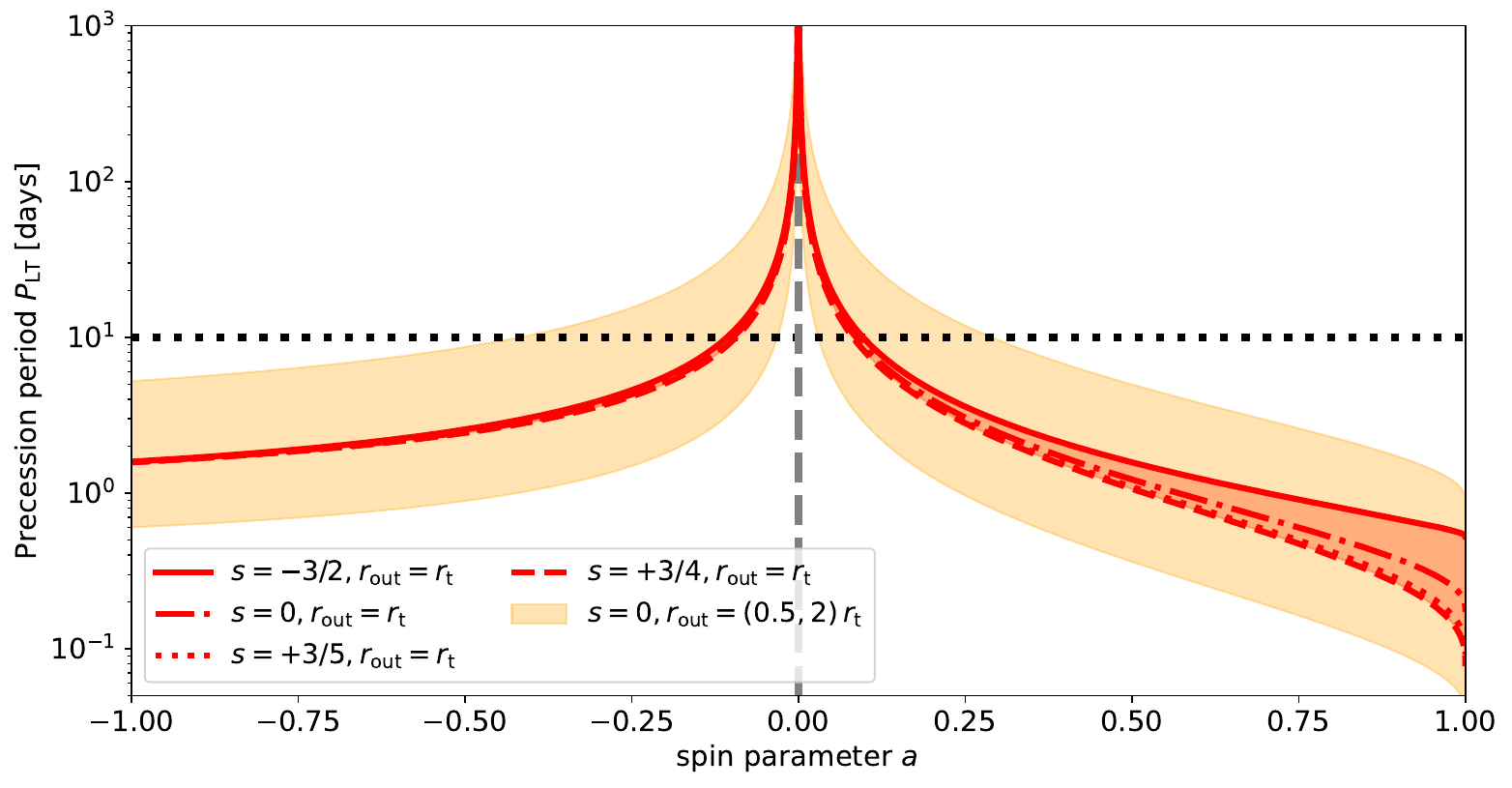}
    \caption{Accretion-disk precession period due to Lense-Thirring effect (in days) as a function of the SMBH spin. We fix the SMBH mass to $M_{\bullet}=10^7\,M_{\odot}$ Different red lines correspond to accretion disks with different surface-density slopes, $s\in(-3/2,0,3/5,3/4)$, $\Sigma \propto r^{-s}$, while the outer radius is fixed to the tidal radius $r_{\rm t}$. The orange-shaded area shows the range of precession periods for the fixed surface-density slope of $s=0$ (constant profile), while the outer radius is varied between $0.5r_{\rm t}$ and $2r_{\rm t}$. The dotted black line depicts the period of 10 days.}
    \label{fig_LT_precession}
\end{figure}

A full TDE occurs when a relatively massive and extended object with the radius $R_{\star}$ and mass $m_{\star}$ (e.g. a star) approaches the SMBH with the mass $M_{\bullet}$ on the scale of its tidal radius or rather below it~\cite{Hills75}. The tidal radius $r_{\rm t}$ expresses the distance scale where the acceleration due to tidal forces of the SMBH acting across the whole stellar body overcomes the gravitational acceleration due to the self-gravity of the star,
\begin{align}
    r_{\rm t}&=\kappa R_{\star}\left(\frac{M_{\bullet}}{m_{\star}} \right)^{1/3} \label{eq_rt}\\
    \frac{r_{\rm t}}{r_{\rm g}}&=\kappa \frac{c^2}{G}R_{\star} M_{\bullet}^{-2/3}m_{\star}^{-1/3}\simeq 10.2\kappa \left(\frac{R_{\star}}{1\,R_{\odot}} \right) \left(\frac{M_{\bullet}}{10^7\,M_{\odot}} \right)^{-2/3} \left(\frac{m_{\star}}{1\,M_{\odot}} \right)^{-1/3}\,,
    \label{eq_rt_rg}
\end{align}
where $\kappa$ is of the order of unity and depends on the radial density profile of the star. Eq.~\eqref{eq_rt_rg} expresses the tidal radius conveniently in gravitational radii, $r_{\rm g}=GM_{\bullet}/c^2$. Using the condition $r_{\rm t}\lesssim r_{\rm g}$, one can infer the minimal SMBH mass, $M_{\bullet}\approx [c^6 R_{\star}^3/(G^3 m_{\star})]^{1/2}\sim 3.2 \times 10^8\,(R_{\star}/1\,R_{\odot})^{3/2} (m_{\star}/1\,M_{\odot})^{-1/2} M_{\odot}$, for which a Solar-type star does not get disrupted but it is instead swallowed as a whole (this order-of-magnitude estimate assumes that the SMBH horizon is located on the scale of a gravitational radius). The TDE is characterized by an increased accretion of a significant fraction of the material from the shredded star, which leads to a luminosity outburst by several orders of magnitude with respect to the quiescent state. Approximately one half of the material becomes ejected from the system, and the other half circularizes to form an accretion disk~\cite{Rees88}.
Theoretical investigation as well as MHD simulations predict that the post-TDE accretion rate evolves with time as a power law, $\propto t^{-5/3}$, and the observed TDEs have shown that this is a viable model for the luminosity evolution as well, especially further away in time from the luminosity peak~\cite{Lodato09, KovacsStermeczky23}.
However, full TDEs are unpredictable and require rapid follow-ups to constrain parameters of the black hole, such as the mass and the spin\cite{2024Natur.630..325P} (likely very small electric charge of SMBHs\cite{2019Obs...139..231Z,2019JPhCS1258a2031Z} appears to be beyond the possibilities of current instruments, with the potential exception of the Galactic center SMBH where weak charge constraints can be set\cite{2018MNRAS.480.4408Z,2020ApJ...897...99T}). 
In doing so we may learn more about the properties of black holes at the population level. More specifically, high-cadence two-band follow-up observations of TDEs are likely to shed light on the following,
\begin{itemize}
    \item the production mechanism of the UV, optical, and the X-ray emission of TDEs, which can be inferred from the time lag analysis of the FUV, NUV, optical, and X-ray light curves. Detailed monitoring of selected targets will help distinguish among several proposed mechanisms, see also Fig.~\ref{fig_TDE_models} for the illustration of shock-at-apocenter model (top left panel), series of discrete self-interactions (top right panel), elliptical disk (bottom left panel), and the reprocessing scenario (bottom right panel)\cite{2015ApJ...806..164P,2017ApJ...837L..30P};
    \item accretion flow state transitions for a sample of TDEs\cite{2021ApJ...912..151W,2023A&A...669A..75L}, especially in synergy with X-ray monitoring. As the relative accretion rate transitions from the one close to the Eddington rate to the sub-Eddington one, the system is expected to transition from high/soft state dominated by the softer UV/optical emission ($<2$ keV) of an accretion disk to low/hard state dominated by the harder ($>2$ keV) non-thermal power-law emission of hot corona and the inner ADAF;
    \item time-resolved spectral energy distributions, including radio, optical, UV, and X-ray data, of jetted TDEs, which correspond to $\sim 1\%$ of all TDEs\cite{2022Natur.612..430A}. This would allow the detailed understanding of the jet formation and collimation shortly after the TDE when the disk is still geometrically thick due to the super-Eddington accretion rate (slim disks). In case the jet is oriented close to the line of sight, the emission is enhanced due to Doppler-beaming\cite{2018MNRAS.478.3199B,2023ApJ...951..106B} and thus can be detected even for high-redshift ($z\gtrsim 1$) sources at cosmological distances;
    \item X-ray and UV quasiperiodic ``flares'' shortly after the TDE, potentially due to the Lense-Thirring (LT) precession of a newly formed accretion disk that is misaligned with respect to the equatorial plane of the SMBH. In combination with the independent constraints on the SMBH mass, accretion-disk surface density slope ($\Sigma(r)\propto r^{-s}$), outer disk radius $r_{\rm o}$, and the measured LT precession period $P_{\rm LT}$, one can constrain the SMBH spin $a_{\bullet}$ using the following formula for the LT precession of a rigid-like disk\cite{2020SSRv..216...35S,2024Natur.630..325P},
    \begin{align}
       P_{\rm LT}(a_{\bullet})&=2\pi \sin{\psi}\left(\frac{J}{N} \right)\,\notag\\ 
       &=\frac{\pi G M_{\bullet}(1+2s)}{c^3 a_{\bullet}(5-2s)}\frac{r_{\rm o}^{5/2-s}r_{\rm i}^{1/2+s}[1-(r_{\rm i}/r_{\rm o})^{5/2-s}]}{1-(r_{\rm i}/r_{\rm o})^{1/2+s}}\,,  
       \label{eq_LT_precession_period}
    \end{align}
    where $\psi$ is the misalignment angle between the accretion disk and the SMBH equatorial plane, $J$ is the total angular momentum of the disk, and $N$ is the integrated LT torque. The inner radius of the accretion disk $r_{\rm i}$ is assumed to be given by the innermost stable circular orbit, and thus it is determined by the SMBH spin. In Fig.~\ref{fig_LT_precession}, we plot the LT precession period in days, see Eq.~\eqref{eq_LT_precession_period}, as a function of the SMBH spin for the case of $M_{\bullet}=10^7\,M_{\odot}$. We see that depending on the SMBH spin, the accretion-disk precession period and thus flux variations can range from $\sim 0.1-1$ day for higher SMBH spins to $\sim 10$ days for lower spins. 
    Rapid follow-up, high-cadence monitoring of TDEs could thus reveal X-ray and UV quasiperiodic variations due to LT precession\cite{2024Natur.630..325P}, which would lead to the constrains on the SMBH spins at different redshifts. Constraining SMBH spin distribution is crucial for the understanding of the evolution of the SMBH growth during the cosmic history (accretion-dominated, merger-dominated or both\cite{2005ApJ...620...69V}); 
    \item late-time UV emission associated with TDEs. Models predict\cite{2020MNRAS.492.5655M,2021MNRAS.504.5144M} that after the initial power-law decrease of the accretion rate and hence also the luminosity, the accretion disk reaches steady-state and dominates the observed UV plateau at late times. Using relativistic accretion-disk models, it is possible to infer the SMBH mass just based on the detection of the UV flux density of this late-time TDE emission.
\end{itemize}

AGN flaring events likely occur when an increased amount of material is accreted (e.g. QPOs by disk tearing~\cite{Musoke23}) or when the accretion disk itself is perturbed by an external object~\cite{Linial24}.
By studying these flaring events with rapid UV observations, certain properties of the accretion disk may be studied.
Some of these properties include the density of the accretion disk and the feedback mechanisms between material brought to the disk by the perturber.

AGN are known to evolve with time, entering varying states, some of which are dormant.
The transition between states is relatively short lived over cosmic time scales.
However, changing-look AGN (CL-AGN) are likely this transitionary period and their study offers us a rare look into this process~\cite{Jana24}.
CL-AGN seem to be associated with the rapid changes in the structure of the innermost accretion flow, e.g. the case of 1ES 1927+654 seems to indicate the destruction of the corona and its recreation on a yearly timescale\cite{2020ApJ...898L...1R,2021ApJS..255....7R}. A possible interpretation of such abrupt changes involves the interaction of the denser circumnuclear gaseous material (e.g. post-TDE streams) with the accretion flow, which initially enhances the accretion rate and subsequently causes the depletion of the inner accretion flow and the destruction of hot corona. After some time, the inner disk is refilled and the corona reforms as well. Supermassive binary black holes shortly before the merger are also candidates for curious and specific variability patterns. While acting as relatively normal AGN before the merger, where the emission originates in the circumbinary disk and minidisks around each black hole, shortly after the merger the X-ray emission nearly disappears while the UV emission originating in the circumbinary disk is largely unaffecred. This could be a tell-tale sign of supermassive binary black holes merging\cite{2023MNRAS.526.5441K}.
\emph{QUVIK} offers the ability to regularly monitor these objects in the UV bands and can help to find and probe these rare transitions.

Other variability phenomena are associated with the changes in the variability pattern rather than the ``look'' itself. Such a special case is NGC 5548, which exhibited the phenomenon known as the ``BLR on holiday", which shows that the BLR emission variability can decouple from the photoionizing continuum variability over $\sim$180 day time scale~\cite{Goad16}.
It was suggested that the ``holiday'' was caused by a dense out-flowing cloud which obscured the underlying continuum, breaking the correlation with the BLR emission~\cite{Dehghanian19}. In fact, the disk winds with a wide range of column densities may be a ubiquitous phenomenon. When they are fully transparent, standard BLR reverberation takes place. When the density increases above the threshold value, it induces the decoupling between the continuum and the BLR variability, which can address ``BLR holidays". For even larger densities over a certain time span, i.e. when the disc wind prevents the continuum emission from reaching BLR clouds at larger distances, broad lines can disappear as seen in some CL-AGN~\cite{Dehghanian19}.

Another relatively rare transient event in strongly lensed AGN~\cite{Wambsganss} is a microlensing high-magnification event (HME), where a micro-caustic from the lensing galaxy tomographically scans the AGN over month-to-year time-scales~\cite{Vernardos23}.
Depending on the lensing configuration, micro-caustics from a lensing galaxy may amplify a region on the order of micro- to nano-arcseconds~\cite{Neira20}.
This is an external effect to the AGN and occurs in each image individually.
Therefore, by monitoring these events and comparing the flux evolution between images across various wavelengths, we have the ability to probe the structure of the AGN's central engine~\cite{Mediavilla15b, Ledvina18, Best24a}.
While the microlensing event occurs over weeks or months, the most interesting period of time is the peak of the HME which only lasts on the timescale of days to weeks.
Recent work has gone into predicting these events from expected data from wide-field surveys~\cite{Fagin24b}.
With the UV emission of the AGN expected to be relatively compact compared to the optical accretion disk or the BLR, \emph{QUVIK} has the ability to rapidly monitor these HMEs as they unfold and can assist in probing the central engine of the AGN.\\
\textit{\underline{QUVIK observational strategy:}} In order to distinguish among different origins of the UV light in TDE sources, see Fig.~\ref{fig_TDE_models}, or to study processes in CL-AGN well-sampled UV, optical, and X-ray light curves need to be compared with model light curves at the corresponding wave-bands. Detailed modeling of such processes is beyond the scope of this overview. Here we just compare relevant timescales that the UV mission should be able to resolve for this purpose, see Table~\ref{tab_BHtimescales}. The shortest timescales of a fraction of the day are dynamical and light-crossing timescales. The thermal timescale of a few days is related to the heating/cooling-front propagation. The longest timescales are related to the perturbation propagation and viscous processes ($\gtrsim $ year). In general, since all of the timescales are proportional to the SMBH mass, they can vary by orders of magnitude. In addition, the viscous timescale, on which accretion processes take place, can be significantly shortened if instead of the whole disc at the distance $r$ only its narrow ring of the width $\Delta r$ is affected by e.g. an instability. Then the viscous timescale is $\tau_{\rm visc}(\Delta r)\approx (\Delta r)/r)\tau_{\rm visc}$ \cite{2020A&A...641A.167S}. Therefore an initial high-cadence monitoring of any nuclear transient at $\sim 0.1$ day can resolve several dynamical processes taking place in the inner accretion disk or affecting its narrower part. From an observational point of view, the duration of the monitoring as well as the cadence will need to be adjusted for intermediate-redshift sources; e.g. for $z\sim 1$, the necessary duration of the campaign would be prolonged by $(1+z)\sim 2$, while the characteristic variability timescale would also get stretched by the same factor. However, for intermediate-redshift sources, \textit{QUVIK} can probe the progressively shorter-wavelength spectral regions towards the FUV domain in the rest frame of the source ($175\,{\rm nm}/(1+z)\sim 87.5\,{\rm nm}$ in the FUV band and $325\,{\rm nm}/(1+z)\sim 162.5\,{\rm nm}$ in the NUV band), for which the variability timescale tends to be smaller, and therefore the monitoring cadence should remain approximately the same to capture it. In order to observe the UV transients at the satisfactory signal-to-noise ratio, such as TDEs, the redshift distribution of the detectable sources peaks towards low-redshift, nearby nuclei. Assuming the SMBH accreting at the Eddington limit shortly after the star is tidally disrupted in its vicinity, the bolometric luminosity is $L_{\rm bol}\sim 1.26 \times 10^{44}\,{\rm erg\,s^{-1}}$ for $M_{\bullet}=10^6\,M_{\odot}$. Most of this luminosity is initially radiated away in the optical/UV domain, hence $L_{\rm UV}\sim 10^{44}\,{\rm erg\,s^{-1}}$ at $\lambda=300$ nm. This yields the apparent AB magnitude of $m_{\rm AB}=-2.5\log{F_{\nu}}-48.60\sim 17.4$ for $z=0.1$, which is well within the \textit{QUVIK} sensitivity range. Adopting the conservative sensitivity limit $m_{\rm AB}\lesssim 21.5$ mag for \textit{QUVIK}, we obtain the limiting redshift of $z\lesssim 0.53$ for the detectability of bright UV transients (TDEs). For this redshift, the corresponding time dilation factor is $\sim 1.5$, while at the same time \textit{QUVIK} will probe the spectral region at $325\,{\rm nm}/(1+z)\sim 217\,{\rm nm}$ in the rest frame of the source.

\begin{table}[tbh!]
    \centering
      \caption{Comparison of relevant timescales for the SMBH accretion disks. $M_{\bullet}$ stands for the SMBH mass, $r$ for the distance from the SMBH, $\alpha$ is a viscous parameter, and $h/r$ is a scale-height-to-radius ratio.}
    \begin{tabular}{c|c}
    \hline
    Time scale & Relation  \\
    \hline
    Dynamical     &  $\tau_{\rm dyn}=12.2\left(\frac{r}{20~r_{\rm g}}\right)^{3/2}\left(\frac{M_{\bullet}}{10^8~M_{\odot}} \right)$~hours  \\ 
    Light-crossing & $\tau_{\rm lc}=2.7\left(\frac{r}{20~r_{\rm g}} \right)\left(\frac{M_{\bullet}}{10^8~M_{\odot}} \right)$~hours \\
    Thermal  & $\tau_{\rm th}=5.1\left(\frac{\alpha}{0.1} \right)^{-1}\left(\frac{r}{20~r_{\rm g}}\right)^{3/2}\left(\frac{M_{\bullet}}{10^8~M_{\odot}} \right)$~days\\
    Perturbation propagation & $\tau_{\rm prop}=509\left(\frac{\alpha}{0.1} \right)^{-1}\left(\frac{h/r}{0.01} \right)^{-1}\left(\frac{r}{20~r_{\rm g}}\right)^{3/2}\left(\frac{M_{\bullet}}{10^8~M_{\odot}}\right)$~days\\
    Viscous & $\tau_{\rm visc}=139\left(\frac{\alpha}{0.1} \right)^{-1}\left(\frac{h/r}{0.01} \right)^{-2}\left(\frac{r}{20~r_{\rm g}}\right)^{3/2}\left(\frac{M_{\bullet}}{10^8~M_{\odot}}\right)$~years\\
    \hline
    \end{tabular}
    \label{tab_BHtimescales}
\end{table}

\subsection{Repeating nuclear transients}
\label{subsec_repeating}

Repeating nuclear transients have been associated with stars or compact remnants orbiting SMBHs, causing recurrent flares on the timescales of several months to hours (see Suková et al.\cite{2024arXiv241104592S} for a review). Repetitive outbursts on longer timescales are likely caused by stripping the outer layers of a star while the inner core remains intact (partial TDEs). This is the case when the relative impact parameter $\beta=r_{\rm t}/r_{\rm p}<\beta_{\rm crit}$, where $r_{\rm t}$ is the tidal radius and $r_{\rm p}$ is the pericenter distance of the star. For main-sequence stars, $\beta_{\rm crit}$ was constrained to be $\sim 2-3$\cite{2020ApJ...904...99R} and for these values, the star likely gets fully disrupted. Partial TDEs exhibit variability in both UV/optical and X-ray wavebands, often with a time delay due to different processes (e.g. AT2018fyk\cite{2023ApJ...942L..33W} or ASASSN0-14ko\cite{2021ApJ...910..125P}).

Currently the shortest repeating nuclear transients associated with galactic nuclei are Quasiperiodic Erupters (QPEs~\cite{2019Natur.573..381M,2021Natur.592..704A}), which exhibit large-amplitude soft X-ray flares on the timescale of hours to days; however, they do not exhibit significant variability in the optical/UV domains; see, however, the analysis of the ``Ansky'' source that shows signs of UV variability\cite{2025arXiv250407169H}. Recently, a class of repeating soft X-ray nuclear transients on the intermediate timescales of tens of days has been revealed (represented by Swift J0230-28\cite{2023NatAs...7.1368E,2024NatAs...8..347G}), which shares most of the spectral properties with QPEs, and they likely correspond to partial TDEs of non-main-sequence stars, such as red giants or even gas-giant planets\cite{2024arXiv241105948P}. In the case of Swift J0230-28, no significant variability in the optical/UV bands was found, though a radio flare was associated with the start of one of the X-ray flares\cite{2024NatAs...8..347G}.

The peak soft X-ray luminosity of QPE outbursts reaches $\sim10^{42}$ erg s$^{-1}$. The exact mechanism for their production is uncertain -- the proposed models span from accretion-disk intrinsic instabilities\cite{2020A&A...641A.167S,2023A&A...672A..19S} to bodies periodically crossing an accretion disk\cite{2021ApJ...917...43S,2024arXiv241104592S}. A favored class of models involve extreme mass-ratio inspirals (EMRIs)\cite{2024MNRAS.532.2143K}, i.e.\ stars or compact remnants which collide with the accretion disks of supermassive black holes (SMBHs), producing flares through disk–star interactions involving shocks within the ejected material \cite{2023ApJ...957...34L}. This class of models can also address the variability in the flare recurrence timescale, in particular the long-short recurrence timescale pattern, which can be attributed to the EMRI prograde orbital (Schwarzschild) precession in combination with the accretion-disk precession \cite{2023ApJ...957...34L,2023A&A...675A.100F}.

Since the EMRI distance can be expressed in gravitational radii as $r_{\rm EMRI}/r_{\rm g} \propto M_{\bullet}^{-2/3}P_{\rm QPE}^{2/3}$, longer-period QPEs are expected to result in softer flares peaking in the UV domain for the fixed SMBH mass\cite{linial2024_UV}. Similarly, for the QPE periodicity comparable to the X-ray QPEs, the flares are expected to be softer and potentially detectable in the UV domain in lower-mass SMBH systems. Also, since some X-ray QPEs seem to be associated with post-TDE nuclei with a decreasing relative accretion rate, the flare spectra are expected to get softer with the drop in the accretion rate, potentially ceasing to be detected in the X-ray domain and appearing in the UV bands \cite{linial2024_UV} when the relative accretion rate reaches $\dot{m}\sim \mathcal{O}(10^{-2})$. The detection of such events in the UV domain is feasible with future missions, such as \textit{ULTRASAT} and \textit{QUVIK}, which have sensitivities down to $\sim22$ AB magnitude. The \textit{QUVIK} mission has the projected sensitivity limit of $\sim 22$ AB mag in the NUV band (325 nm), which implies the monochromatic flux density limit of $F_{\nu}\sim 5.75 \times 10^{-29}\,{\rm erg\,s^{-1}\,cm^{-2}\,Hz^{-1}}$ or $\nu F_{\nu}\sim 5.31\times 10^{-14}\,{\rm erg\,s^{-1}\,cm^{-2}}$ in the NUV band. UV QPEs in galaxies could thus be detectable within a distance of up to $d_{\rm L}=235 (L_{\rm QPE}/3.5\times 10^{41}\,{\rm erg\,s^{-1}})^{1/2} (\nu F_{\nu}/5.31\times 10^{-14}\,{\rm erg\,s^{-1}\,cm^{-2}})^{-1/2}\,{\rm Mpc}$ ($z=0.053$), assuming the typical UV QPE luminosity of $3.5 \times 10^{41}\,{\rm erg\,s^{-1}}$\cite{linial2024_UV}. The detection and monitoring of UV QPEs would significantly advance our understanding of repeating nuclear transients, their potential association with EMRIs~\cite{2024MNRAS.532.2143K,2024arXiv241104592S}, and the evolution of SMBH accretion disks plausibly following a TDE\cite{2024Natur.634..804N}.\\
\textit{\underline{QUVIK observational strategy:}} With the high-cadence monitoring of selected candidates, \textit{QUVIK} can detect quasiperiodicity at the level of $\sim 0.2$ days or about 5 hours. This way, the repeating UV nuclear transients can be extended towards significantly shorter periods. It will be necessary to conduct longer monitoring for a few months so that the periodic signal can effectively be disentangled from stochastic variability. The nature of the suggested UV variability of QPE sources\cite{2025arXiv250407169H} can thus be better constrained. In combination with the X-ray data, it will be possible to test the origin of UV flares, specifically if they originate due to the reprocessing of harder radiation or if they are produced in shocks due to star-disk interactions or alternatively due to accretion-disk instabilities. The sample of detectable repeating nuclear UV transients is located towards lower redshifts. When we adopt the redshift distribution of detected 13 X-ray QPEs~\cite{2024arXiv241104592S}, we obtain the median redshift of $z=0.024$ with the 16\% and the 84\% percentiles of $0.018$ and $0.091$, respectively (the mean redshift is $0.047 \pm 0.036$, while the minimum and the maximum redshifts are 0.0151 and 0.1177, respectively). For the peak UV luminosity of $L_{\rm 300}\sim 10^{42}\,{\rm erg\,s^{-1}}$ at 300 nm (NUV band)\cite{linial2024_UV}, we obtain the AB magnitude $m_{\rm AB}=-2.5\log{F_{\nu}}-48.60\sim 22.4$ for $z=0.1$. Taking the conservative sensitivity limit of $m_{\rm AB}\lesssim 21.5$, we infer the redshift limit of $z\lesssim 0.065$ for the detection of QPE-like UV transients, which is consistent with the estimate above considering the uncertainties of peak UV luminosities and of the sensitivity limit. A new class of (repeating) UV transients can also be detected in connection with blazar jets and related $\gamma$-ray flares\cite{2024MNRAS.535.2742B}. Interactions of stars orbiting the SMBH with the nuclear outflow and the jet could result in recurring shocks\cite{2016MNRAS.455.1257Z,2020ApJ...903..140Z,2020A&A...644A.105H,2025MNRAS.tmp..765K} with the emission spectrum peaking at progressively lower energies as the shocked material cools downstream. The UV counterparts of the blazar sources could be searched for within the dedicated observational program aimed at radio-loud AGN sources.

\section{Synergy of QUVIK observations with other missions}
\label{sect: Discussion}

The \emph{QUVIK} observing program concerning AGN and other sources will consist of two main strategies:
\begin{itemize}
\item[(i)] the telescope will perform dedicated monitoring of the variability of selected bright AGN ($m_{\rm AB}\lesssim 18$ mag) located at low to intermediate redshifts ($z\lesssim 1$). These AGN programs and campaigns will be selected based on a competitive peer-review process following annual proposal calls available to the international community. After the established proprietary period, the data will be freely available via the archive portal.
\item[(ii)] the satellite will conduct high-cadence follow-up observations ($\sim 0.1-1$ day cadence) of target-of-opportunity (ToO) transient sources identified by observing survey programs with much larger fields of view, such as Vera C. Rubin's Legacy Survey of Space and Time (LSST) \cite{2019ApJ...873..111I}, the Ultraviolet Transient Astronomy Satellite (\emph{ULTRASAT}) \cite{2024ApJ...964...74S}, or the Zwicky Transient Facility (ZTF) \cite{2019PASP..131a8002B}. \textit{QUVIK} will also carry as a secondary payload the GRB localizing instrument \textit{GALI}\cite{2020SPIE11444E..6ER}, which will also serve as a trigger for $\gamma$-ray transients that can be followed up in the UV domain. This may also be relevant for $\gamma$-ray flares in radio-loud AGN, in particular those whose jets point nearly at us (blazars or flat-spectrum radio quasars) and thus are Doppler-boosted and can be detected at cosmological distances.
\end{itemize}   

\begin{figure}
    \centering
    \includegraphics[width=\textwidth]{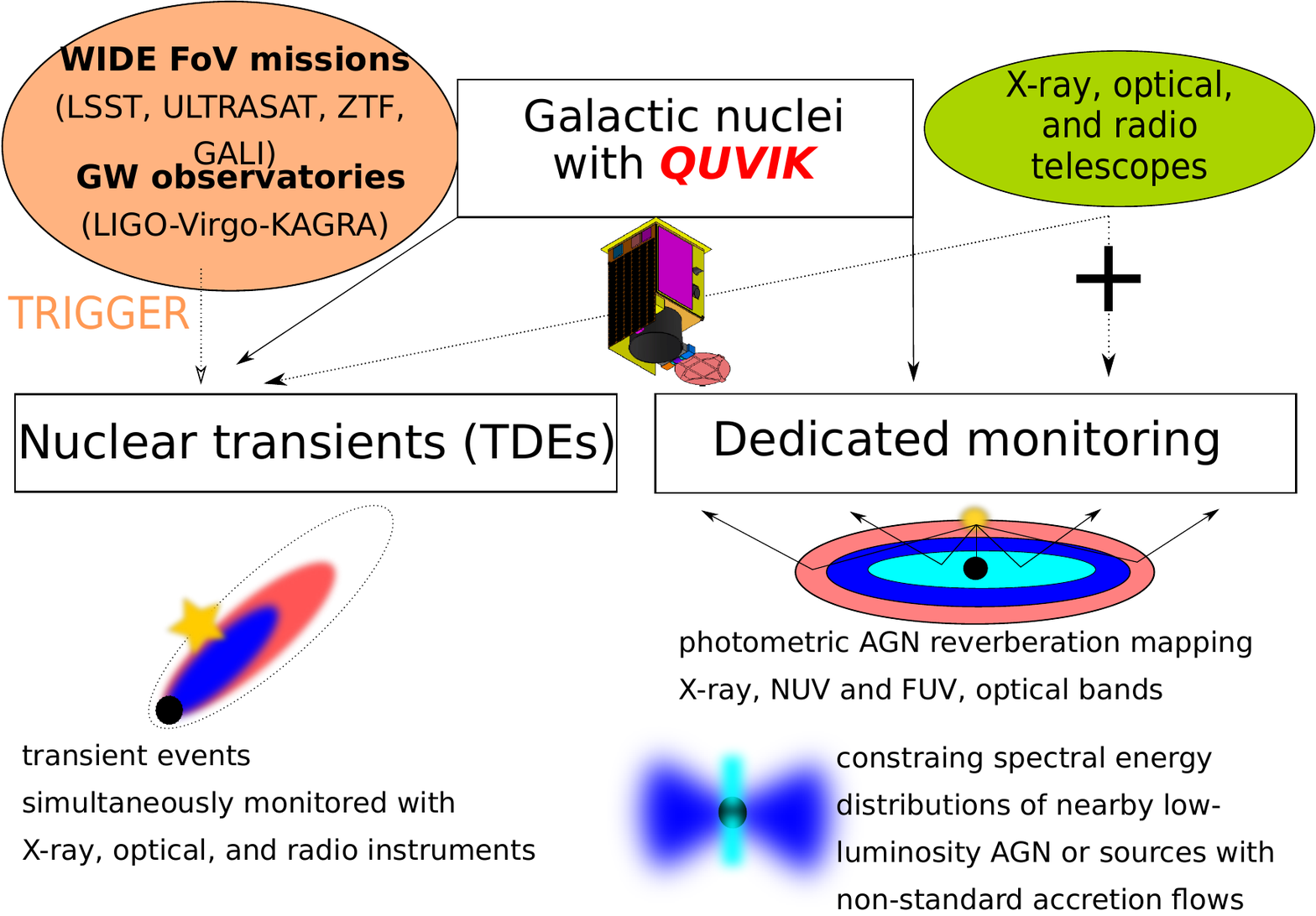}
    \caption{A scheme of the \textit{QUVIK} observing program focused on AGN and transient nuclear sources. The program will consist of a high-cadence follow-up of nuclear transients, such as TDEs, and of a dedicated two-band monitoring (NUV and FUV bands) of selected bright AGN, potentially in synergy with other multi-wavelength and multi-messenger facilities.}
    \label{fig_obs_program}
\end{figure}

We illustrate the \emph{QUVIK} observing program in synergy with other multiwavelength and multi-messenger observatories in Fig.~\ref{fig_obs_program}. The legacy of \emph{QUVIK} observations will also be beneficial for the cross-calibration with the future UV missions, such as the \emph{UltraViolet EXplorer} (UVEX\cite{2021arXiv211115608K}), which is planned to be launched at the beginning of 2030s and will observe the entire sky in the NUV and FUV bands.  

\section{Summary}
\label{sect: Summary}

\textit{QUVIK} is an approved versatile two-band UV photometry space mission currently under development. It is designed to perform observations of the early UV emission of kilonovae, which is its primary scientific objective. In addition, it will conduct monitoring and follow-up observations of various sources, including stellar transients, hot stars, binary stars, stellar associations and clusters. Building on the legacy of previous successful UV space telescopes, \textit{QUVIK} will also bring valuable results concerning active galactic nuclei (AGN) and nuclear transients. The main areas of AGN research include:
\begin{itemize}
    \item reverberation mapping of accretion disks. \textit{QUVIK} can conduct high-cadence observations of AGN variability simultaneously in both FUV and NUV bands. For the sources with the SMBH mass of $10^9\,M_{\odot}$, the required cadence is $\lesssim 1$ day, while for $10^8\,M_{\odot}$, the cadence needs to increase to $\sim 0.1$ days, while the campaign duration should last for about half a year,
    \item the monitoring duration will need to be adjusted taking into account the redshift of the nuclear source; however, because of the sensitivity limits in the NUV and the FUV bands, the focus will be on relatively low-redshift sources with $z\lesssim 0.5$, for which the time dilation factor is $(1+z)\lesssim 1.5$. The cadence will remain to be $\lesssim 1$ day since for higher-redshift sources the UV telescope probes progressively shorter-wavelength spectral regions that vary on shorter timescales in the rest frame,
    \item \textit{QUVIK} can perform early follow-up of tidal disruption events (TDEs), with the focus on detecting their early UV emission, which will help to constrain its origin. In addition, with the detection of late-UV emission of post-TDE accretion disks as well as quasiperiodic variabilities associated with Lense-Thirring precession, \textit{QUVIK} can significantly contribute to the determination of masses and spins of several SMBHs,
    \item \textit{QUVIK} will contribute to the discoveries of repeated nuclear transients and their monitoring, including repeated partial TDEs and plausible UV quasiperiodic erupters. 
\end{itemize}
\textit{QUVIK} will operate in synergy with wide field-of-view observing facilities, including ULTRASAT, Vera C. Rubin Legacy Survey of Space and Time as well as with gravitational-wave observatories (LIGO-Virgo-KAGRA). \textit{QUVIK}, with its agile mode of operation and fast response, will also provide important synergies for the larger future UVEX mission equipped with a 2-degree-long, multi-width UV spectrometer across a wide spectral band (115-265 nm, $R>1000$) in addition to a two-band FUV-NUV imager~\cite{Kulkarni2021}.  Importantly, its potential detections of repeating nuclear transients, some of which may be associated with extreme- and intermediate-mass ratio inspirals, will be beneficial for the upcoming space-borne low-frequency gravitational-wave observatories (LISA) in terms of constraining their cosmological volume densities and looking for potential electromagnetic counterparts of gravitational-wave sources~\cite{2024MNRAS.532.2143K}.



\subsection* {Data availability statement}

All the data used and obtained in this manuscript will be made available upon request.

\subsection* {Disclosure statement}

The authors declare there are no financial interests, commercial affiliations, or other potential conflicts of interest that have influenced the objectivity of this research or the writing of this paper.

\subsection* {Acknowledgments}
MZ, HB, MP, VK, MM, and PS acknowledge the support from the Czech Science Foundation (GACR) Junior Star grant no. GM24-10599M. NW, L\v{S}, and ILG were supported by the Czech Science Foundation (GACR) grant GX21-13491X. PK and ML acknowledge the continuing support of the OPUS-LAP/GAČR-LA
bilateral research grant (2021/43/I/ST9/01352/OPUS22 and
GF23-04053L). JL acknowledges the support from the J.W. Fulbright Commission in the Czech Republic. J\v{R} acknowledges the support from the Czech Science Foundation (GA\v{C}R) grant no. GC24-11487J.




\vspace{2ex}\noindent\textbf{Michal Zaja\v{c}ek}, the lead author of this text and description, is a researcher at the Masaryk University in Brno, Czechia, and is currently a member of the \textit{QUVIK} science team and the leader of its AGN/nuclear transients working group. He received his BSc and MSc degrees from Charles University in Prague and a PhD degree in experimental physics from the University of Cologne in the framework of the International Max Planck Research School for Astronomy \& Astrophysics in Bonn. Currently he focuses on the analysis of the processes in the Galactic center, active galactic nuclei, and nuclear transients, with the focus on recurrent phenomena associated with bodies orbiting supermassive black holes.

\vspace{1ex}
\noindent Biographies and photographs of the other authors are not available.

\listoffigures

\listoftables

\end{document}